\documentclass[aps,pra,twocolumn,showpacs,floatfix]{revtex4}

\usepackage{amsmath}
\usepackage{amssymb}
\usepackage{graphicx}
\usepackage{epstopdf}
\usepackage{verbatim}
\usepackage{color}
\usepackage{appendix}
\usepackage[colorlinks, linkcolor=blue, citecolor=blue, urlcolor=blue,breaklinks=true]{hyperref}
\usepackage{soul}
\usepackage{dsfont}

\graphicspath{{figures/}}

\newcommand{\be}{\begin{equation}}
\newcommand{\ee}{\end{equation}}
\newcommand{\bsubeq}{\begin{subequations}}
\newcommand{\esubeq}{\end{subequations}}

\def\ud{\mathrm{d}}
\newcommand{\bmx}{\begin{pmatrix}}
\newcommand{\emx}{\end{pmatrix}}
\newcommand{\bsmx}{\begin{smallmatrix}}
\newcommand{\esmx}{\end{smallmatrix}}
\newcommand{\vect}[2]{\begin{pmatrix} {#1} \\ {#2} \end{pmatrix}}

\begin{document}

\title{Critical exponent of quantum phase transitions driven by colored noise}
\author{D. Nagy}
\author{P. Domokos}

\affiliation{Institute for Solid State Physics and Optics, Wigner Research Centre, Hungarian Academy of Sciences, H-1525 Budapest P.O. Box 49, Hungary}
\begin{abstract}

We demonstrate that criticality in a driven-dissipative system is
strongly influenced by the spectral properties of the reservoir. We
study the open-system realization of the Dicke model, where a bosonic
cavity mode couples to a large spin formed by two motional modes of an
atomic Bose-Einstein condensate. The cavity mode is driven by a high
frequency laser and it decays to a Markovian bath, while the atomic
mode interacts with a colored reservoir. We reveal that the soft mode
fails to describe the characteristics of the criticality. We calculate
the critical exponent of the superradiant phase transition and
identify an inherent relation to the low-frequency spectral density
function of the colored bath. We show that a finite temperature of the
coloured reservoir does not modify qualitatively this dependence on
the spectral density function.

\end{abstract}

\pacs{05.30.Rt,42.50.Pq,37.10.Vz,03.75.Kk} 

\maketitle

\section{Introduction}

Quantum phase transitions in driven-dissipative systems opened up a
novel research area in the field of critical phenomena
\cite{DallaTorre2010Quantum,Diehl2010Dynamical,DallaTorre2012Dynamics,Chitra2015Dynamical,Schiro2016Exotic}. These
transitions lie beyond the standard classification of classical
dynamical or equilibrium phase transitions, and define completely new
universality classes \cite{Sieberer2013Dynamical,Marino2016Driven,Marino2016Quantum,Lang2016Critical}.
In an open quantum system, the critical behaviour appears in the state
formed by the dynamical equilibrium of the external driving and
dissipation processes. The abrupt symmetry breaking change of such a
steady state takes place when the external control parameters are
continuously tuned across the critical point
\cite{LeBoite2013SteadyState,LeBoite2014BoseHubbard,Griesser2013Lightinduced,Ostermann2016Spontaneous}. The correlation
functions at the critical point are determined by nonequilibrium noise
rather than thermal or ground-state quantum fluctuations
\cite{Sachdev2011Quantum,Strack2011Dicke,Piazza2014Umklapp,Hwang2015Quantum,Niederle2016Ultracold}.

Recent cavity quantum electrodynamics (cavity QED) experiments with
ultracold atoms strongly coupled to the radiation field of a
high-finesse resonator led to the observation of a variety of quantum critical
phenomena
\cite{Baumann2010Dicke,Baumann2011Exploring,Mottl2012RotonType,Schmidt2014Dynamical,Baden2014Realization,Klinder2015Dynamical,Klinder2015Observation,Landig2016Quantum,Kollar2015Adjustablelength,Kollar2016SupermodeDensityWavePolariton}. 
Owing to the high degree of precision in
controlling the system parameters in these experiments, the critical
regime can be explored with sufficient resolution, such that
quantitative measurements of the critical exponent are available \cite{Brennecke2013Realtime}. Remarkably, the theoretical description of the system can be significantly simplified to treating the dynamics of only a few relevant modes of the interacting many-body system. Therefore, many-atom cavity QED systems offer a suitable platform to describe generic features of quantum criticality.

The effective Hamiltonian underlying the observed superradiant phase
transition in Refs.~\cite{Baumann2010Dicke,Klinder2015Dynamical} is
closely related to the Dicke-model which describes the interaction of
a single bosonic mode with a large spin-N degree of freedom. In the
thermodynamic limit, $N\rightarrow \infty$, the Dicke-model is exactly
solved by the corresponding mean-field model \cite{Vidal2007Finitesize}. Approaching the critical
point, the incoherent population in the bosonic mode diverges
following a power law with exponent $1/2$ \cite{Nagy2010DickeModel}. Dissipation can be taken
into account in the mean-field theory which amounts to a model of
coupled boson modes with linearized fluctuations. This is still a
solvable model which results in a critical exponent 1
\cite{Nagy2011Critical}. The experiment, which is intrinsically a
dissipative system due to the presence of photon losses from the
cavity, confirmed that the value of exponent is indeed in the vicinity
of 1 \cite{Brennecke2013Realtime}. It was argued that the dissipative
system, although the actual temperature is $T=0$, is a classical
system and the critical point corresponds to a thermal phase
transition at a certain effective temperature
\cite{Torre2013Keldysh}. The exponent 1 observed experimentally is in
accordance with this claim. On the other hand, one may note that
different bosonic modes are subject to reservoirs at different
effective temperatures, i.e., there is no global effective
temperature. Eluding the classical analogy, too, the bosonic
modes exhibit a certain amount of entanglement even in the lossy
system. Moreover, the logarithmic negativity, used as an entanglement
  measure, undergoes a non-analytic behaviour in the critical point 
\cite{Nagy2011Critical}.  

In a recent letter, we studied how the characteristics of the
dissipation determine the critical exponent
\cite{Nagy2015Nonequilibrium}. We introduced a simple model based on
two coupled boson modes which corresponds well to the open-system
Dicke model describing effectively the relevant dynamics of the
experimental system in the normal phase, i.e., below the critical
point. One of the modes representing the cavity mode is lossy simply
due to the photon leakage from the cavity, which is a Markovian
process to very high accuracy. Besides this simple relaxation process,
we took into account that the other bosonic mode representing the
density-wave mode of the ultracold atom cloud relevant to the criticality is also damped. The principal loss mechanism is the collision induced scattering into the continuum of density-wave excitation
modes. The damping process is a kind of Beliaev damping of
phonons dressed by photons  \cite{Konya2014Photonic,Konya2014Damping}. It is a complicated theoretical task 
to describe this effect at a microscopic level. For
simplicity, we modelled the dissipation of the atomic boson mode phenomenologically,
by assuming that it is subjected to a colored
reservoir with sub-ohmic spectral density function. Within this
approach, we pointed out the existence of a relation between the
critical exponent and the exponent of the power-law spectral density
function of the reservoir. As a consequence, the critical exponent can
take on, in principle, any value in an interval, including values
below 1.  In the present paper we generalize our calculation into two
important directions. First, we extend the class of sub-ohmic
reservoirs to the super-ohmic ones, i.e., we treat in full generality
the spectral density function. Second, we assume finite temperature of
this reservoir in order to check if the unusual exponent values
survive thermal effects at finite temperature.

The paper is structured as follows. In Sect.~2, we recall the main steps to derive the effective two-boson model used for studying the dynamics in the vicinity of the critical point. We define the colored reservoir on a microscopic basis, and introduce the Keldysh path integral of the action for the lossy bosonic modes.  In Sect.~3, we consider the critical phenomenon by using the various components of the Keldysh Green's function. First, we discuss the singularities of the retarded Green's function, and then we study the spectral properties of both the photonic and the atomic modes by means of the Keldysh component of the corresponding reduced Green's functions. The critical exponent is extracted numerically from the divergence of the correlation functions. The finite temperature effects are calculated within the same formalism in Sect.~4. Finally, we conclude in Sect.~5.

\section{Model}

\subsection{The Dicke Hamiltonian and its bosonization}

Ultracold atom experiments where a laser-driven Bose-Einstein
condensate is dispersively coupled to a high-finesse optical cavity
realize the driven Dicke model \cite{Nagy2010DickeModel}, that is
\begin{equation}
\label{eq:H_t}
H = \omega_a\,a^\dagger{}a + \omega_b\, \hat S_z  + y (a\,
e^{i\omega_pt} + a^\dagger e^{-i\omega_pt}) \frac{\hat S_x}{\sqrt{N}}\; .
\end{equation}
The large spin of $N/2$ represents the motional state of $N$ atoms
distributed in two different motional modes of the ultracold gas \cite{Nagy2010DickeModel}. The
laser pump incident on the atoms is far detuned from an atomic
transition, while it is quasi-resonant with a single cavity mode.
Hence, the atoms mediate effective photon scattering between the laser
and the cavity mode, that is accompanied by atomic recoil. This scattering
process gives rise to the time dependent interaction term in
Eq.~(\ref{eq:H_t}) \cite{Ritsch2013Cold}.

In a frame rotating at the driving frequency $\omega_p$, the Hamiltonian becomes time independent, 
\begin{equation}
H = \delta_a\,a^\dagger{}a + \omega_b\, \hat S_z  + y (a + a^\dagger)
\frac{\hat S_x}{\sqrt{N}}\, ,
\end{equation}
where the detuning parameter $\delta_a = \omega_a - \omega_p$ expresses the effective photon energy in the harmonically driven system. Since the spin represents the
collective motional state of the atoms, the corresponding frequency
given by the recoil frequency ($\omega_b = \omega_R$) is the smallest
frequency scale in the system ($\omega_R/\omega_p \leq 10^{-10}$).

The superradiant phase transition takes place in the thermodynamic
limit ($N \rightarrow\infty$) of the Dicke model
\cite{Garraway2011Dicke}. When reaching a critical coupling $y_c^{\rm
  GS} = \sqrt{\delta_a\omega_b}$, a coherent photon field amplitude 
builds up spontaneously in the cavity and the spin becomes polarized along the $S_x$
direction. Due to the global coupling of the model (all the spins are  
coupled to the same bosonic mode, i.e., the coordination number is infinite)
the phase transition can be described within mean-field theory, which
is {\it exact} in the thermodynamic limit \cite{Cardy1996Scaling}.
In order to develop the mean-field theory, one introduces the
Holstein-Primakoff boson representation of the collective spin $\hat
S_-=\sqrt{N- b^\dagger b}\, b$, $\hat S_+=b^\dagger\, \sqrt{N-
  b^\dagger b}$ and $\hat S_z= b^\dagger b- N/2$. The Hilbert space of
the boson operator $b$ is truncated for $b^\dagger b \geq N$. Then the
operators are displaced by their mean values $a \rightarrow \alpha +
a,\; b \rightarrow \beta + b$ and the square roots in the Hamiltonian are expanded up to
second order in the operators $a$, $b$, and their adjoints (the corrections are of the order of
$1/N$). The mean fields are readily solved in
Ref.~\cite{Emary2003Chaos} and with dissipation in
Ref.~\cite{Nagy2011Critical}.  In the normal phase $\alpha=\beta=0$,
while in the superradiant phase the mean fields are nonzero: $\alpha,
\beta \propto \sqrt{N}$.

Our central goal is to study the dissipative effects on the quantum
critical point of the model. We can freely choose to approach the critical point of
the system from the side of the normal phase ($y<y_c$), where the
Dicke problem reduces to the following simple Hamiltonian of two
coupled bosonic modes,
\begin{equation}
\label{eq:dicke}
H_S/\hbar = \delta_a\, a^\dagger{}a + \omega_b\, b^\dagger{}b 
+ \frac{y}{2}(a + a^\dagger)(b + b^\dagger)\,.
\end{equation}

We shall consider the interplay between two dissipation channels.  
Mode $a$ is assumed to be a cavity mode driven by a
laser and emitting ``high frequency'' photons into the vacuum, thus
its decay is unaffected by the interaction between modes $a$ and
$b$. Mode $b$ couples, on the other hand, to a colored reservoir that is sampled
at the eigenfrequencies of the coupled system described by the above Hamiltonian. 
We apply the Keldysh path integral approach  to
calculate the dissipative effects of the two reservoirs \cite{Torre2013Keldysh}, which is a suitable tool to
obtain directly the steady-state properties of the system.

\subsection{Markovian dissipation of the cavity mode}

For mode $a$ the
photon loss dissipation can be described in frequency space by the
action
\begin{equation}
\label{eq:Sa}
S_a = \int \frac{\ud\omega}{2\pi}\left(a_{\rm cl}^*, a_{\rm
  q}^*\right) \bmx 0 & \omega-\delta_a-i\kappa
\\ \omega-\delta_a + i\kappa & 2i\kappa \emx \vect{a_{\rm
    cl}}{a_{\rm q}}\,,
\end{equation}
where $a_{\rm cl}(\omega)$ and $a_{\rm q}(\omega)$ are the classical
and quantum fields that correspond to the operator $a$.  Action
(\ref{eq:Sa}) describes the standard Markovian decay process of the cavity
mode by means of a constant $\kappa$. The high-frequency drive ensures that the cavity dissipation is
unaffected by the interaction between modes $a$
and $b$, since $y \ll \omega_p$ even in the superradiant phase. The
parameter $\delta_a=\omega_a-\omega_p$ expresses the fact that the
cavity mode frequency is referenced to the driving frequency
$\omega_p$. 

\subsection{General description of the dissipation of the atomic mode}

Let us consider a bosonic bath composed of modes $c_k$ interacting
with mode $b$ according to the Hamiltonian
\begin{equation}
H_{bc} = \sum_k \omega_k c^\dagger_k{}c_k + \sum_k g_k(b^\dagger{}c_k
+ c^\dagger_k{}b)\,.
\end{equation}
Integrating out the reservoir modes $c_k$ in the Keldysh path integral
formalism, we obtain the following dissipative action
for the oscillator $b$,
\begin{equation}
\label{eq:Seff_b}
S_{b} = \int_\omega 
\left(b_{\rm cl}^*, b_{\rm q}^*\right)
\bmx
 0 & \omega-\omega_b - K^A \\
\omega-\omega_b - K^R & D
\emx
\vect{b_{\rm cl}}{b_{\rm q}}\,,
\end{equation}
where $b_{\rm cl}(\omega)$ and $b_{\rm q}(\omega)$ are the classical
and quantum fields corresponding to mode $b$, 
with self energies
\bsubeq
\label{eq:K_D}
\be
K^{R/A}(\omega) = \sum_k \frac{g_k^2}{\omega-\omega_k \pm i\eta}
\ee
that incorporate the frequency shift and damping originating from the
coupling to the reservoir. Note that this is sometimes referred to as 
the level-shift function in quantum optics
\cite{Kurucz2010Multilevel,CohenTannoudji1992AtomPhoton}. Finally, the Keldysh component of the action is
\be
\label{eq:G_K_b}
D(\omega) = 2\pi{}i\sum_k g_k^2F(\omega)\delta(\omega-\omega_k)\,, \ee
\esubeq with the function $F(\omega) =
\coth\left\{(\hbar\omega-\mu)/k_BT \right\}$, which takes
into account the thermal occupation of the bath modes at a given
frequency. For generality, we enable a chemical potential $\mu$ for
the reservoir which is needed for the low temperature limit of the bosonic modes of the reservoir.
We consider the thermal effects only in mode $b$, since the high-frequency mode $a$ is still
effectively at zero temperature, $\hbar \omega_b \sim k_B T \ll
\hbar\omega_p$.

\subsection{Colored reservoir}
The genuine ingredient of the system is that the reservoir coupled to the
low-frequency mode $b$ is colored. Such a non-Markovian bath arises from
quasi-particle scattering in the Bose-Einstein condensate, {\it e.g.}, 
from a Beliaev-type damping process \cite{Konya2014Photonic,Konya2014Damping}. Instead of providing a
microscopic theory, we consider a general class of reservoirs described by the
coupling-density profile 
\begin{equation}
\rho(\omega) =
\sum_kg_k^2\delta(\omega-\omega_k)
\end{equation}
The essential point is that this bath is sampled at the
eigenfrequencies of the coupled system.  Since we focus on the
soft-mode criticality of the phase transition, the relevant property
of the bath is its low-frequency behaviour.  We use the classification
of the colored reservoirs introduced by Leggett
\cite{Leggett1987Dynamics}, and we assume a power-law frequency
dependence of $\rho(\omega)$ with exponent $0 < s < 2$, \be
\label{eq:rho}
\rho(\omega) =
\gamma\Theta(\omega)\,\frac{\left(\omega/\omega_b\right)^s}{1 +
  \left(\omega/\omega_M\right)^4}\, 
\ee 
where $\Theta(\omega)$ is the Heaviside function, $\gamma$ is the
dissipation strength and $\omega_M$ is a cutoff frequency. The cutoff
at the fourth power in the denominator enables us to consider both the
sub-Ohmic ($s<1$) and the super-Ohmic ($s>1$) cases.  Our results will be, of course, independent of the the cutoff in the end. The range $s>2$ can be easily described in the same way and gives no new physics. 

The coupling-density function determines completely  the level-shift function Eq.~(\ref{eq:K_D}a), which can be expressed by using complex analytic extension. When  $\rho(\omega)$ is an analytic function of the real variable $\omega$,
the level-shift function can be extended to the complex plane analytically 
\be
\label{eq:Kz}
K(z) = \int_0^\infty\frac{\rho(\omega)}{z-\omega}\,\ud\omega\,,
\ee
such that the retarded $K^R(\omega)$ is extended analytically to the
upper half plane (${\rm Im}z > 0$), and the advanced $K^A(\omega)$ is
extended to the lower half plane (${\rm Im}z < 0$). The reverse expression reads
\be
K^{R/A}(\omega) = \lim_{\eta\rightarrow{}0}K(\omega\pm i\eta)\,.
\ee
To uncover the properties of the complex function $K(z)$, we evoke the 
relation
\be
\lim_{\eta\rightarrow{}0}\frac{1}{x\pm i\eta} = {\cal P}\frac{1}{x}
\mp i\pi\delta(x)\,.
\ee
Substituting it into the integral in Eq.~(\ref{eq:Kz}), we obtain
\be 
\label{eq:sokhotski}
\lim_{\eta\rightarrow{}0}K(\omega\pm i\eta) = 
{\cal P}\int_0^\infty\frac{\rho(\omega^{'})}{\omega-\omega^{'}}
 \,\ud\omega^{'} \mp i\pi\rho(\omega) \,,
\ee
where the real part describes the frequency shift and the imaginary
part yields the damping rate. As $\rho(\omega)$ is nonzero only for
$\omega > 0$, $K(z)$ has a branch cut along the positive real axis,
and its imaginary part vanishes along the negative real axis.  At this
point our investigation differs from the standard textbook example,
where $K^{R/A}(\omega)$ is evaluated at a large positive frequency.
Usually, in the frame of the Markov approximation, the lower boundary of
the integral in Eq.~(\ref{eq:Kz}) is extended to $-\infty$, which would
continue the branch cut of $K(z)$ to the whole real axis. On the
contrary, we are looking for the effects of dissipation around zero
frequency, hence we continue analytically
$K^{R/A}(\omega)$ to the complex plane by keeping the lower limit of the integral.

\begin{figure}[htb]
\includegraphics[width=0.6\columnwidth]{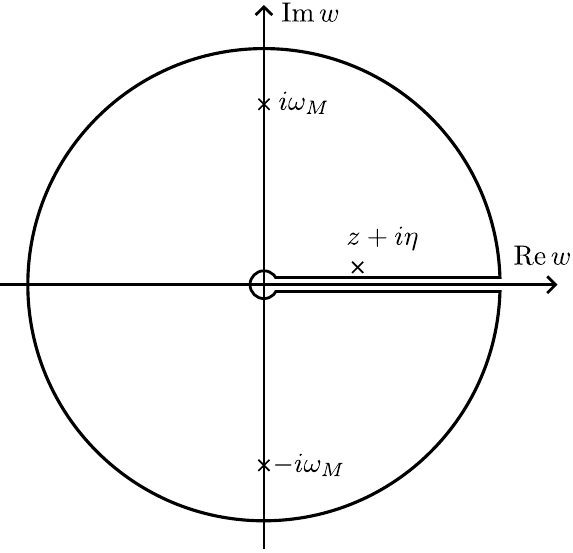}
\caption{Contour for calculating the integral. It starts in the
  infinitesimal vicinity of the origin and goes along the real axis, 
  then makes a full circle in the infinity. It comes back along the 
  real axis, to be finally closed by the circle surrounding the
  origin, where the fractional power function is non-analytic. 
  The symbol $\times$ indicates the poles. 
}
\label{fig:contour}
\end{figure}

With the coupling-density profile given in Eq.~(\ref{eq:rho}), it is
possible to calculate $K^{R}(z)$ by evaluating the integral
\be
\label{eq:K_Rz}
K^R(z) = \int_0^\infty\frac{\rho(\omega)}{z-\omega +i\eta}\,\ud\omega\,,
\ee
by complex contour integration. First, we extend the function of $\omega$ in the integrand
analytically to the complex plane. Since $K^{R}(z)$ has a branch cut
along the positive real axis, we choose the complex function $(-z)^s$
with $s<1$, and calculate the following integral on the closed contour 
given in Fig.~\ref{fig:contour},
\be
\label{eq:contour_integral}
R(z) = \oint_C
\frac{(-w/\omega_b)^s}{(z-w+i\eta)[(w/\omega_M)^2+1]}\ud{}w\,, 
\ee
where $z$ and $w$ are complex numbers. Note that we assume $s<1$ for
simplicity, when $s>1$ one should separate the fractional part of $s$
in the numerator as $(-w/\omega_b)^{[s]} (-w/\omega_b)^{\{s\}}$, and
recombine these functions after the integration. The radius of the
small circle around $w=0$ is finally made to approach zero, while the
radius of the large circle is taken to infinity, that
eliminates their contributions to the integral. The two integrals
performed in opposite directions along the positive real axis give
different results, since they are on opposite sides of the branch cut:
above and below the positive real axis, the phase of the function $(-w)^s$
is $e^{-is\pi}$ and $e^{is\pi}$, respectively. After exchanging the
limits of the integral that goes on the line below the branch cut, we
obtain 
\be
\label{eq:K_z_trick}
R(z) = K^R(z)[e^{-is\pi} - e^{is\pi}]/\gamma\,.  
\ee 
On the other hand, the contour integral Eq.~(\ref{eq:contour_integral}) 
can be calculated by means of the residue theorem. The integrand has 
three poles at $w=z+i\eta$ and $w=\pm i\omega_M$. All of them are
inside the contour, resulting
\begin{multline}
R(z) = 2\pi{}i\Bigg[-\frac{(-(z+i\eta)/\omega_b)^s}{
    \big(\frac{z+i\eta}{\omega_M}\big)^2 + 1}  +
  {\rm Re}
  \bigg\{\frac{(i\omega_M/\omega_b)^s}{1-i\frac{z+i\eta}{\omega_M}}
  \bigg\} \Bigg] \,.
\end{multline}

The second term increases with the cutoff frequency $\omega_M$, and it
gives a large contribution to the real part of $K^R(z)$, hence
renormalizing the frequency $\omega_b$. To eliminate the effects of
the cutoff, we do the standard renormalization by taking the limit
$\omega_M\rightarrow\infty$ and incorporating the diverging real part
into $\omega_b$. In this way, the first term results
\begin{equation}
R(z) = -2\pi{}i\left(-\frac{z+i\eta}{\omega_b}\right)^s .
\end{equation}
Substituting this result into the left hand side of
Eq.~(\ref{eq:K_z_trick}), we obtain 
\be
\label{eq:K_R_z}
K^{R}(z) =
\frac{\pi\gamma}{\sin\,s\pi}\left(-\frac{z+i\eta}{\omega_b}\right)^s.
\ee 
We keep $i\eta$ in the expression to mark that the calculation is
not valid when the pole of the integrand is on the positive real
axis. Normally, when calculating the level-shift function for real
$\omega$, it is the $\pm i\eta$ which distinguishes between the
retarded and advanced Green's functions having the symmetry
\mbox{$K^A(\omega) = [K^R(\omega)]^*$}.  Here, in the $\eta
\rightarrow 0$ limit,  we obtain 
\be
\label{eq:K_R_A}
K^{R/A}(\omega) = \frac{\gamma\pi}{\sin\,s\pi}
(\omega/\omega_b)^s[\Theta(\omega) e^{\mp is\pi} + \Theta(-\omega)]\,.
\ee 
The Heaviside-function $\Theta(\omega)$ assures the property
stemming from Eq.~(\ref{eq:sokhotski}) that the level-shift function
is real for $\omega<0$. The above result is valid up to $s<c$, where
$c=2$ is the exponent of the cutoff. Note that the trigonometric factors
are periodic in $s$, namely the  $K^R(\omega>0)$ part is invariant to 
$s\rightarrow s +1$, and $K^R(\omega<0)$ is invariant to $s\rightarrow
s +2$.

In order to complete the calculation of the elements of the action in Eq.~(\ref{eq:Seff_b}), we note that the Keldysh component Eq.~(\ref{eq:K_D}b) is simply expressed as 
\be
\label{eq:D}
D(\omega) = 2i\pi{} \rho(\omega) \,.  
\ee

\section{Non-equilibrium phase transition}

The driven Dicke model exhibits a non-equilibrium phase transition 
that we analyzed previously in Ref.~\cite{Nagy2011Critical} without
the colored bath. We demonstrated that the cavity photon loss
fundamentally changes the critical properties of the system with
respect to the closed Hamiltonian model. The second order phase
transition takes place in the steady state, where the critical
exponent is changed to $1$ (instead of the mean-field exponent $1/2$
found in the ground state). Also, the critical point is shifted
upward by the cavity decay rate. In the following, we describe how the colored reservoir modifies the
critical point of the open system. 

\subsection{The full Keldysh action}

The Keldysh action corresponding to the  interaction term in the Hamiltonian, in Eq.~(\ref{eq:dicke}),
reads
\begin{equation}
\label{eq:Sab}
S_{ab} = -\frac{y}{2}\int_\omega
\left[(a_q + a_q^*)(b_{cl} + b_{cl}^*) + (a_{cl} + a_{cl}^*)(b_q + b_q^*)\right]\,,
\end{equation}
expressed with the classical and quantum components. The
counterrotating terms in the Hamiltonian results in terms like
$a_{cl}b_q$ or $a_q^* b_{cl}^*$. Hence, in order to cast the Keldysh
action into the standard matrix form, one doubles the variable space
by introducing fields with negative frequency 
\be {\bf v}(\omega) = \bmx a_{cl}(\omega)
\\ a_{cl}^*(-\omega) \\ b_{cl}(\omega) \\ b_{cl}^*(-\omega)
\\ a_{q}(\omega) \\ a_{q}^*(-\omega) \\ b_{q}(\omega)
\\ b_{q}^*(-\omega) \emx\,. \ee 
Then the total Keldysh action
\be S = S_a + S_b + S_{ab} \,\ee can be written as
\begin{equation}
\label{eq:action_total}
S = \frac12 \int \frac{\ud\omega}{2\pi}
{\bf v}^\dagger
\bmx
 0 & [{\bf G}^A_{4\times{}4}]^{-1}(\omega) \\
[{\bf G}^R_{4\times{}4}]^{-1}(\omega) & {\bf D}^K_{4\times{}4}(\omega)
\emx
{\bf v}\,.
\end{equation}
The $4\times{}4$ blocks are matrix Green's functions:
\begin{multline}
\label{eq:G_R_1_4x4}
[{\bf G}^R_{4\times{}4}]^{-1}(\omega) = \\
\bmx
 [G_a^R]^{-1}(\omega) & 0 & -y/2 & -y/2 \\
 0 & [G_a^A]^{-1}(-\omega) & -y/2 & -y/2 \\
 -y/2 & -y/2 & [G_b^R]^{-1}(\omega)  & 0 \\
 -y/2 & -y/2 & 0 & [G_b^A]^{-1}(-\omega)
\emx\,,
\end{multline}
with the inverse retarded Green's functions of each mode
\bsubeq
\label{eq:G_R_1a}
\be
[G_a^R]^{-1}(\omega) = \omega - \delta_a + i\kappa \,,
\ee
\be
[G_b^R]^{-1}(\omega) = \omega - \omega_b - K^R(\omega)\,.
\ee
\esubeq
Note that the advanced and retarded Green's functions are hermitian adjoint, \mbox{$[G^A_{4\times{}4}]^{-1}(\omega) =
  [[G^R_{4\times{}4}]^{-1}(\omega)]^\dagger$}, whereas the scalars \mbox{$[G^A]^{-1}(\omega) =
[[G^R]^{-1}(\omega)]^*$} are complex conjugate. The inverse of
the Keldysh component describing the fluctuations is a diagonal matrix
\be
{\bf D}^K_{4\times{}4}(\omega) = \mbox{diag}(2i\kappa, 2i\kappa, D(\omega), D(-\omega))\,,
\ee
whith $D(\omega)$ is given in Eq.~(\ref{eq:D}).

\subsection{The soft-mode pole}
\label{subs:poles}
Second order quantum criticality is accompanied by the vanishing of
the frequency of one of the excitation modes, i.e., the soft mode.  In
a dissipative Markovian system the soft-mode frequency is complex, and
the critical properties are encoded in its imaginary part, the
so-called adiabatic decay rate \cite{Kessler2012Dissipative}.
Although this is not the case when the system is non-Markovian, we
first calculate the characteristic frequencies of the system. These
are identified by the poles of the retarded Green's function continued
analytically to the lower half of the complex plane. Note, however,
that one cannot attribute a single complex frequency to describe the
dynamics of the soft mode {because it does not correspond simply to a Lorentzian resonance}. Moreover, the non-Markovian character of
the system is crucially enhanced close to the critical point.

Let us continue analytically the level shift function to the second
Riemann sheet
\be
\label{eq:Kz2}
K^{R}_{\rm II}(z) = \gamma\frac{\pi
  e^{-is\pi}}{\sin\,s\pi}(z/\omega_0)^s\,, 
\ee 
then use the
symmetry property $K^A_{\rm II}(z) = \left[K^R_{\rm
    II}(z)\right]^*$, and finally replace the argument $-\omega$ of the
advanced components with $-z^*$, which ensures the convergence of the
Fourier integrals. 
The poles are defined by the characteristic equation 
 \mbox{${\rm det}[{\bf G}^R_{4\times{}4}]^{-1}(z) = 0$}, i.e., 
\begin{multline}
\label{eq:poles}
[(z+i\kappa)^2-\delta_a^2][(z-i\Gamma(z))^2
  -(\omega_b+\Delta(z))^2] \\ 
 -y^2\delta_a(\omega_b+\Delta(z)) = 0\,, 
\end{multline}
where $\Gamma(z) = (K^R_{\rm II}(z) - K^A_{\rm II}(-z^*))/(2i)$ and $\Delta(z) = (K^R_{\rm II}(z) + K^A_{\rm II}(-z^*))/2$.

\begin{figure}[htb]
\includegraphics[angle=0,width=0.81\columnwidth]{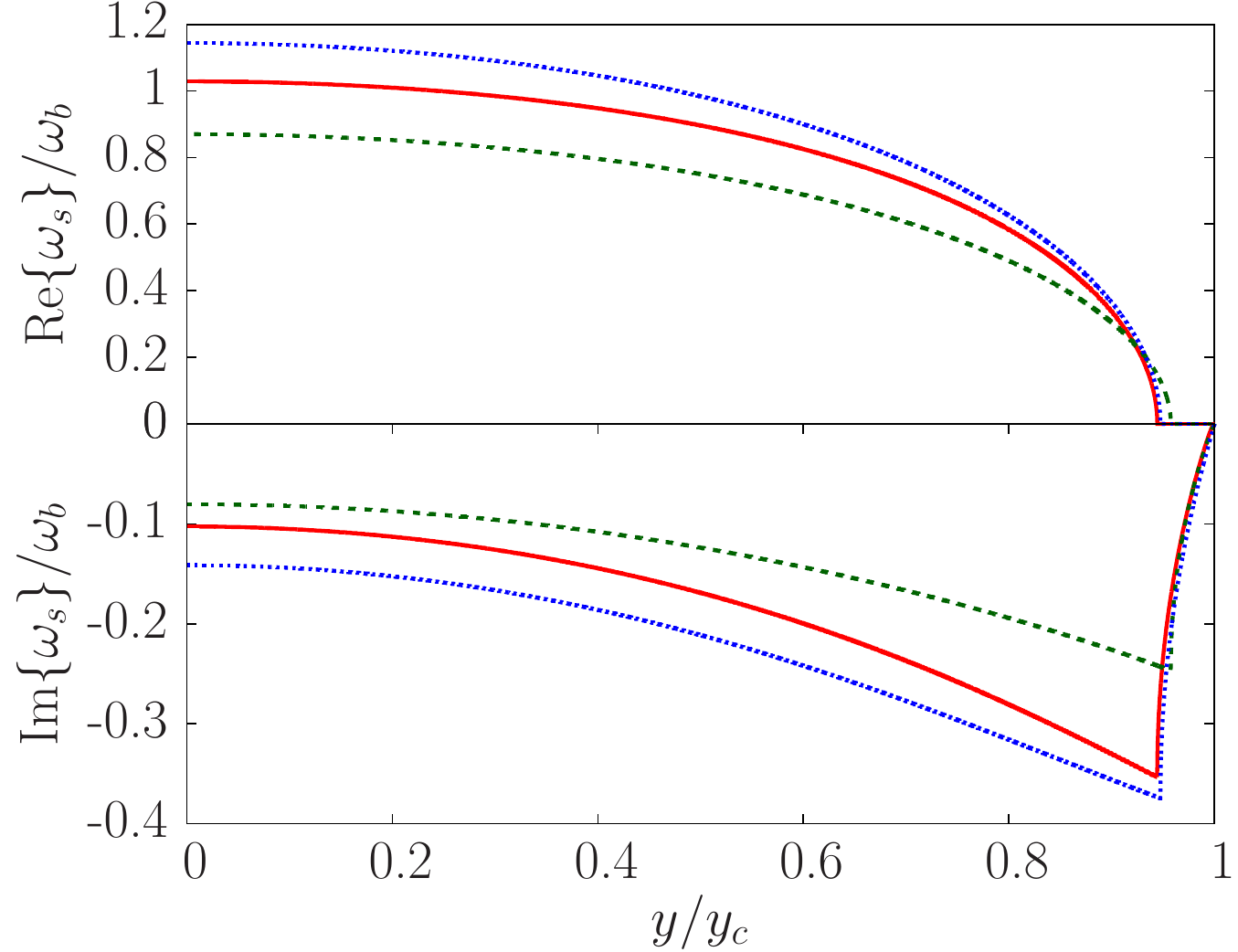}
\caption{(Color online) Real and imaginary parts of the soft-mode
  frequency as a function of the coupling $y$. The parameters are
  $s=0.4$ (solid red) $0.8$ (dashed green), $1.2$ (dotted blue),
  $\mu=0$, $\gamma = 0.1\omega_b$ and $\delta_a=\kappa=2\omega_b$.}
\label{fig:softmode}
\end{figure}
Due to the colored reservoir the 'decay' $\Gamma(z)$ and the 'frequency shift'
$\Delta(z)$ depend explicitly on the complex frequency.  As a result
Eq.~(\ref{eq:poles}) becomes an implicit equation for the
characteristic frequencies. According to the symmetry property
$K^A_{\rm II}(z) = \left[K^R_{\rm II}(z)\right]^*$, the poles come in
pairs, if $z_1$ is a pole, then $z_2 = -z_1^*$ is also a pole of the
retarded photon Green's function. This means that the poles are purely
imaginary or they come in pairs with opposite real part but the same
imaginary part.

Fig.~\ref{fig:softmode} shows how the soft mode frequency solution of
Eq.~(\ref{eq:poles}) vanishes at the critical point as a function of
the control parameter $y$. We plot the soft mode for three different
bath exponents $s=0.4$ (solid red),	$0.8$ (dashed green) and $1.2$
(dotted blue).  At $y=0$ and for $\delta_a > \omega_b$ the soft mode
corresponds to the bare mode $b$, hence the real and imaginary parts
show the frequency shift and decay of mode $b$ due to the interaction
with the colored reservoir. As increasing the coupling parameter $y$, {the dressing by photons leads to 
 mode softening until} reaching a threshold coupling
(around $y\approx 0.95 y_c$). At this point, a linewidth bifurcation takes place
\cite{Eleuch2013Width}. We plot only the upper branch of the imaginary
part, because this is the one associated with to the  soft mode. The critical point is
that where the imaginary part reaches zero. It proves to be
independent of $\gamma$,
\begin{equation}
y_c = \sqrt{\frac{\delta_a^2 + \kappa^2}{\delta_a}\omega_b}\,.
\end{equation}
It is important to note that the colored noise dissipation does not modify the position
of the critical point whereas the dissipation of mode $a$ with $\kappa$ does increase $y_c$. The reason is that 
both the decay rate and the light shift originating from the colored reservoir tend to zero at the critical point,
$\Gamma(z\rightarrow 0) = 0$ and $\Delta(z\rightarrow 0) = 0$, {where the reservoir modes are sampled at zero frequency}. 

\subsection{Correlation functions}

The soft mode pole calculated from Eq.~(\ref{eq:poles}) provides a
good picture far from the critical point, nevertheless it fails to
describe the critical dynamics and exponent.  
These latter can be extracted from the
correlation functions which are provided for by the Keldysh component of the Green's functions.

\subsubsection{Photon correlation function}

We calculate the properties of the photon field by integrating out the
 low-frequency mode $b$ from the total Keldysh action in
Eq.~(\ref{eq:action_total}).  This gives the photon-only action
\be
\label{eq:S_eff}
S_a^{\rm eff} = \int \frac{\ud\omega}{2\pi}
{\bf v}_{a}^\dagger 
\bmx
 0 & [{\bf G}^A_{2\times{}2}]^{-1}(\omega) \\
[{\bf G}^R_{2\times{}2}]^{-1}(\omega) & {\bf D}^K_{2\times{}2}
\emx
{\bf v}_{a}\,,
\ee
where the vector ${\bf v}_{a}$ contains the classical and quantum
components of the photon field
\be
{\bf v}_{a}^\dagger(\omega) = [
a_{cl}^*(\omega), 
 a_{cl}(-\omega), 
a_{q}^*(\omega),
a_{q}(-\omega)] \,,
\ee
and the $2\times 2$ photon Green's functions incorporate the effects
of interaction between the modes. With the integration over mode $b$,
the original inverse retarded photon Green's function transforms as 
$[{\bf G}^R_{2\times{}2,a}]^{-1} \rightarrow   
[{\bf G}^R_{2\times{}2}]^{-1} = [{\bf G}^R_{2\times{}2, a}]^{-1} - 
\frac{y^2}{4}\mathds{I}_{2\times{}2}{\bf G}^R_{2\times{}2,
  b}\mathds{I}_{2\times{}2}$ with the matrix 
$ \mathds{I}_{2\times{}2} =  \left(\bsmx  1 & 1 \\ 1 & 1
\esmx\right)$, by which we obtain
\begin{multline}
\label{eq:G_R_1}
[{\bf G}^R_{2\times{}2}]^{-1} = \\
\bmx
 [G_a^R]^{-1}(\omega) + \Sigma^R(\omega) & \Sigma^R(\omega) \\
 \Sigma^R(\omega) & [G_a^A]^{-1}(-\omega) + \Sigma^R(\omega)
\emx\,.
\end{multline}
The photon self-energy
\be
\Sigma^R(\omega) = -\frac{y^2}{4}\left[\frac{1}{\omega-\omega_b-K^R(\omega)} + \frac{1}{-\omega-\omega_b-K^A(-\omega)}\right]\,
\ee
has the symmetry propery $\left[\Sigma^R(-\omega)\right]^* =
\Sigma^R(\omega)$, which we exploited in Eq.~(\ref{eq:G_R_1}).

The correlation function of the cavity field can be derived from the
$2\times 2$ inverse Keldysh Green's function ${\bf D}^K_{2\times 2}$
of the effective photon-only action (\ref{eq:S_eff}). It is
obtained similarly to the retarded photon Green's function. When
integrating out the fields corresponding to mode $b$, the
original inverse Keldysh component ${\bf D}^K_{2\times{}2,a}(\omega) =
\mbox{diag}(2i\kappa, 2i\kappa)$ transforms as ${\bf
  D}^K_{2\times{}2,a} \rightarrow {\bf D}^K_{2\times{}2} = {\bf
  D}^K_{2\times{}2, a} - \frac{y^2}{4}\mathds{I}_{2\times{}2}{\bf
  G}^K_{2\times{}2,b}\mathds{I}_{2\times{}2}$, with the matrix 
$ \mathds{I}_{2\times{}2} =  \left(\bsmx  1 & 1 \\ 1 & 1
\esmx\right)$. The new inverse Keldysh component is no longer diagonal,
\be
\label{eq:D_K}
{\bf D}^K_{2\times{}2} =
\bmx
 2i\kappa + d(\omega) & d(\omega) \\
 d(\omega) & 2i\kappa + d(\omega)
\emx\,,
\ee
with 
\be
d(\omega) =
-\frac{y^2}{4}\left[G^K_b(\omega) + G^K_b(-\omega)\right]\,,
\ee
where $G^K_b$ is the scalar Keldysh Green's function of mode $b$, that reads 
\be
G^K_b(\omega) = \frac{-D(\omega)}{(\omega-\omega_b -
  K(\omega))(\omega - \omega_b - K^{*}(\omega))}\,. 
\ee
The Keldysh Green's function of the photon field is expressed from
Eq.~(\ref{eq:S_eff}) as ${\bf G}^K_{2\times{}2}(\omega) = -{\bf
  G}^R_{2\times{}2}(\omega){\bf D}^K_{2\times{}2}(\omega){\bf
  G}^A_{2\times{}2}(\omega)$, where the retarded Green's function, the
inverse of Eq.~(\ref{eq:G_R_1}), is
\begin{multline}
\label{eq:G_R}
{\bf G}^R_{2\times{}2} = \frac{1}{{\rm det}[{\bf G}^R_{2\times{}2}]^{-1}}\\
\bmx
 [G_a^A]^{-1}(-\omega) + \Sigma^R(\omega) & -\Sigma^R(\omega) \\
 -\Sigma^R(\omega) & [G_a^R]^{-1}(\omega)+ \Sigma^R(\omega)
\emx\,.
\end{multline}
The advanced Green's function is calculated from
the symmetry ${\bf G}^A_{2\times{}2} = [{\bf G}^R_{2\times{}2}]^\dagger$, 
and ${\bf D}^K_{2\times{}2}$ is given by Eq.~(\ref{eq:D_K}).

\begin{figure}[t]
 \hspace{3mm}\includegraphics[angle=0,width=0.79\columnwidth]{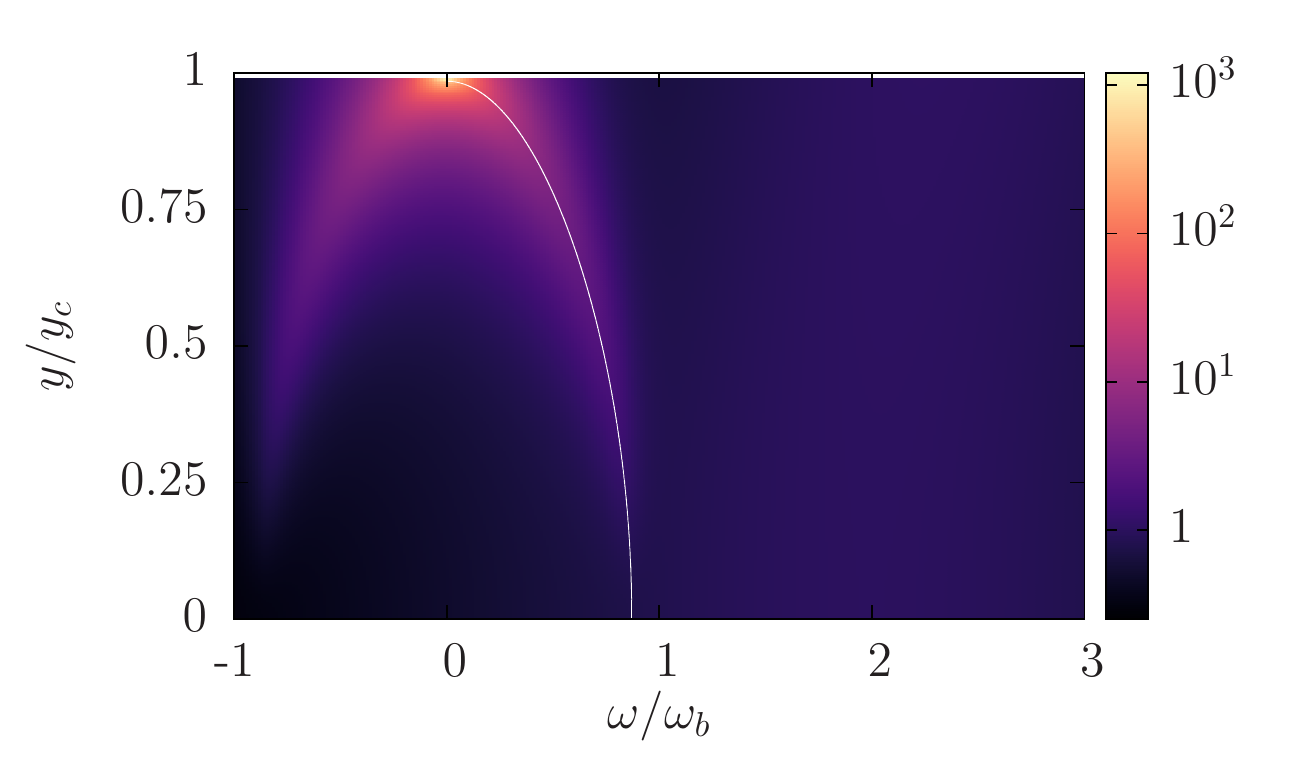}\\[3mm]
  \includegraphics[angle=0,width=0.71\columnwidth]{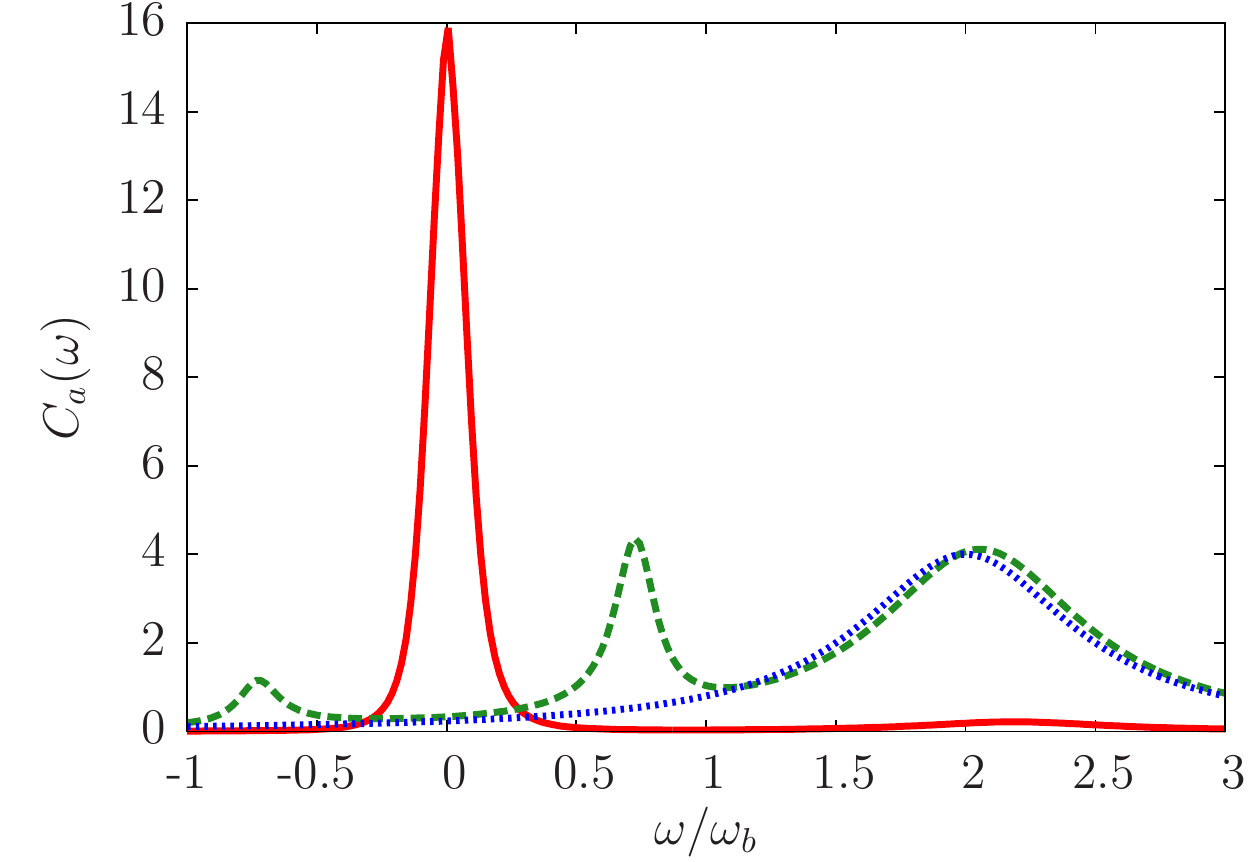}
\caption{(Color online) Correlation function of mode $a$ for various
  coupling strengths approaching the critical point. The top panel
  shows the correlation function in frequency space (x-axis) as a
  function of the relative coupling strength. The white line plots the
  mode softening ${\rm Re}\,\omega_s$ shown in Fig.~\ref{fig:softmode}.
    On the bottom panel we plot the correlation function for three
    given values of the coulping $y=0$ (dotted blue), $0.5$ (dashed
    green) and $0.99y_c$ (solid reed). Parameters:
    $\delta_a=2\omega_b$, $\kappa=0.5\omega_b$, $\gamma=0.1$, $\mu=0$,
    $s=0.8$. The peaks for $y=0.99y_c$ are divided by a factor of
    $20$.}
\label{fig:correl_func}
\end{figure}

The relevant part of the Keldysh Green's function is its first
diagonal element which gives the Fourier transform of the photon
correlation function $C_a(t) = \langle \{a(t), a^\dagger(0)\}\rangle$,
that is $C_a(\omega) = [{\bf G}^K_{2\times{}2}]_{11}$, 
\begin{multline}
\label{eq:correl_func}
C_a(\omega) = \frac{-i}{|{\rm det}[{\bf G}^R_{2\times{}2}]^{-1}|^2} \Big( 2i\kappa \big[ |[G_a^R]^{-1}(-\omega)|^2 \\
+2{\rm Re}\left\{\Sigma^R(\omega)[G_a^R]^{-1}(-\omega)\right\} +
  2|\Sigma^R(\omega)|^2\big] \\ + d(\omega)|[G_a^R]^{-1}(-\omega)|^2 \Big)\,.
\end{multline}
This correlation function corresponds directly to the measurable power
spectrum of the outcoupled cavity field. In
Fig.~\ref{fig:correl_func}, we plot the spectrum $C_a(\omega)$ for
variable coupling $y$ from 0 to the critical point $y_c$ in the form
of a color map. For better visibility, the bottom panel shows the
spectrum at three different values of the coupling $y$, such that the
functions correspond to the cross sections of the color map at zero
coupling $y=0$, at an intermediate coupling ($y/y_c=0.5$) and at a
coupling in the critical domain ($y/y_c=0.99$).  For an uncoupled
cavity mode ($y=0$), the photon correlation function is a Lorentzian
peak of width $2\kappa$ centered at $\omega = \delta_a$. This peak is
hardly visible in the colour code map in the top panel. The Lorentzian
peak extends to negative frequencies, as the whole frequency range is
shifted by the driving frequency in the rotating frame. The negative
frequency part {makes thus sense because of the external driving and}
it gives rise effectively to quantum noise which heats the system into
a non-trivial steady state. For an intermediate coupling strength
($y=0.5 y_c$), a doublet forms symmetrical to $\omega=0$ on the side
of the Lorentzian peak. These two peaks correspond to the two poles
$z_1=-z_2^*$ of the retarded Green's function on the negative complex
half plane discussed in Subsection~\ref{subs:poles}. This
correspondence applies only away from the critical point, where the
dynamics can still be well described by Lorentzian resonances
associated with the poles. {Nevertheless, the soft mode frequencies
  taken from} Fig.~\ref{fig:softmode}, and plotted by a white solid
line on top of the color map, fit accurately to the position of the
peaks of the correlation function. Close to the critical point, in the
interval where the real part of the soft-mode frequency vanishes, the
doublet peaks merge into a large central peak around $\omega=0$ (solid
red line in the bottom panel). In this range the use of the soft mode
picture relying on Lorentzian resonances obviously breaks down.  The
criticality manifests itself in the divergence of this central peak as
the coupling strength converges to $y_c$. In order to obtain the
exponent, this divergence has to be considered and fitted by a power
law.

\subsubsection{Atom correlation function}

The correlation function of the low frequency mode $b$ is derived from
the atom-only action following a similar procedure as described in the
previous subsection. The Keldysh component of the atom-only action reads

\begin{figure}[t!]
 \hspace{3mm}\includegraphics[angle=0,width=0.8\columnwidth]{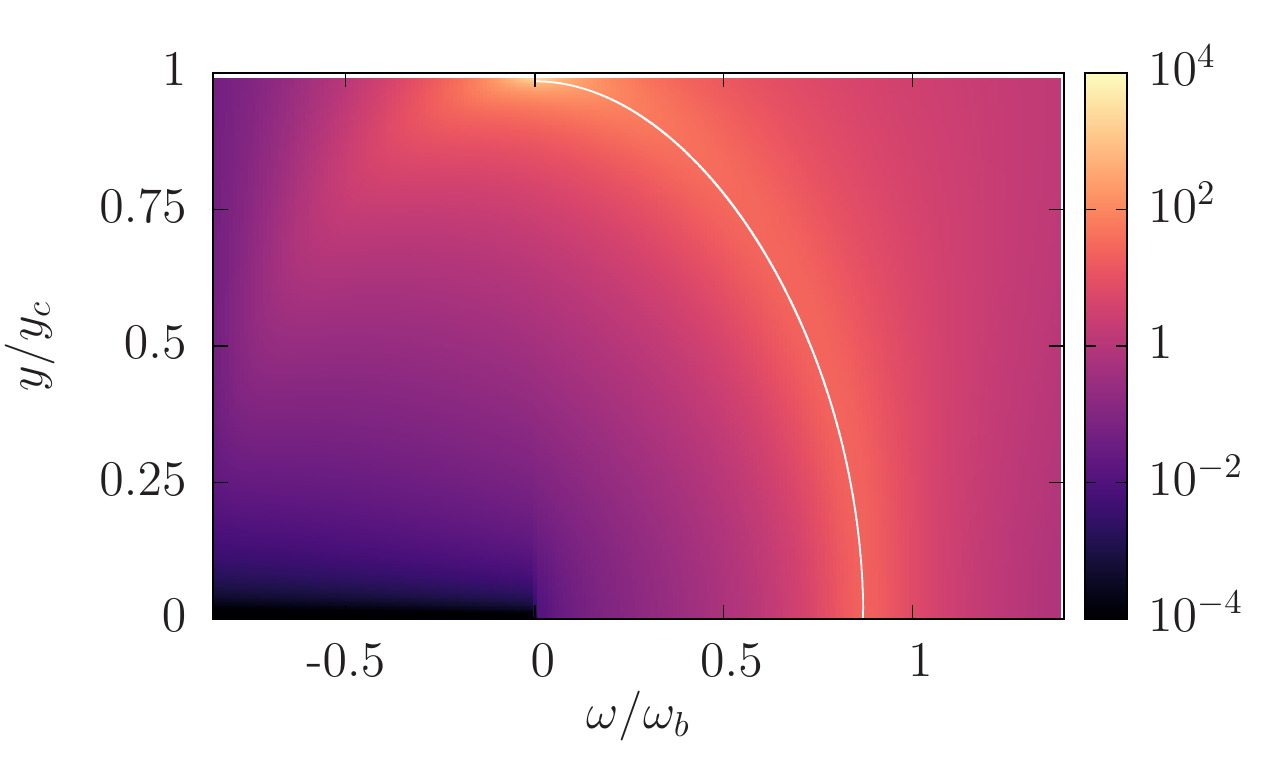}\\[3mm]
  \includegraphics[angle=0,width=0.71\columnwidth]{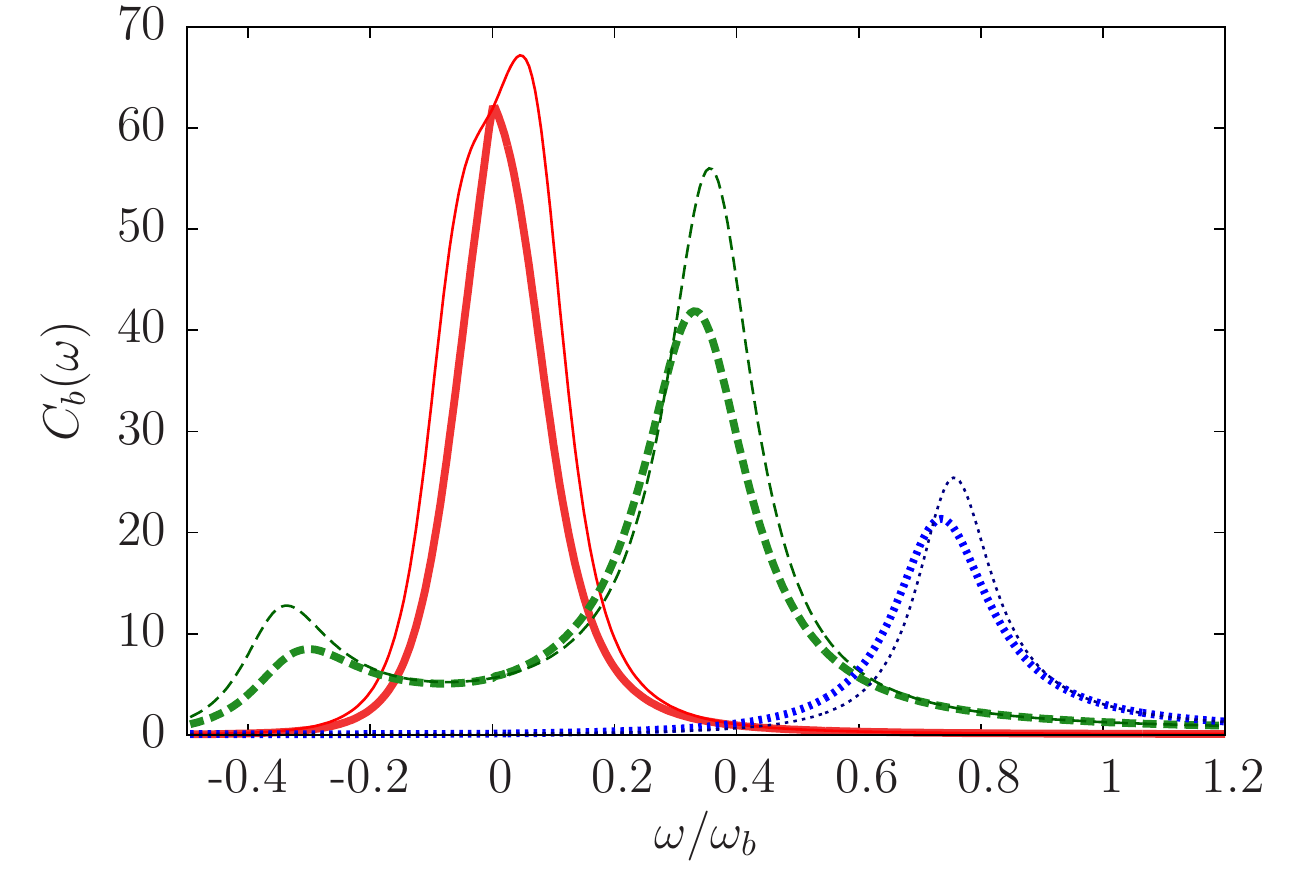}
\caption{(Color online) Correlation function of mode $b$ for various
  coupling strengths approaching the critical point. The top panel
  presents a color map of the correlation function in frequency space
  (x-axis) as a function of the relative coupling strenth (y-axis) for
  $s = 0.8$. The white line shows the mode softening ${\rm
    Re}\,\omega_s$ shown in Fig.~\ref{fig:softmode}. In the bottom
  panel, we compare two complementer bath exponents $s = 0.8$ (thick
  lines) and $s=1.8$ (thin lines) for couplings $y=0$ (dotted blue), $0.5$
  (dashed green) and $0.99$ (solid red). Other parameters are the same as
  in Fig.~\ref{fig:correl_func}.}
\label{fig:correl_func_b}
\end{figure}

\be
\label{eq:D_Kb}
{\bf D}^K_{b,2\times{}2} =
\bmx
 D(\omega) + g(\omega) & g(\omega) \\
 g(\omega) & D(-\omega) + g(\omega)
\emx\,,
\ee
with 
\be
g(\omega) =
-\frac{y^2}{4}\left[G^K_a(\omega) + G^K_a(-\omega)\right]\,,
\ee
where $G^K_a$ is the scalar Keldysh Green's function of mode $a$, 
\be
G^K_a(\omega) = \frac{-2i\kappa}{(\omega-\delta_a)^2 + \kappa^2}\,. 
\ee
The retarded Green's function of the atom field is 
\begin{multline}
\label{eq:G_Rb}
{\bf G}^R_{b,2\times{}2} = \frac{1}{{\rm det}[{\bf G}^R_{b,2\times{}2}]^{-1}}\\
\bmx
 [G_b^A]^{-1}(-\omega) + \Sigma_b^R(\omega) & -\Sigma_b^R(\omega) \\
 -\Sigma_b^R(\omega) & [G_b^R]^{-1}(\omega)+ \Sigma_b^R(\omega)
\emx\,,
\end{multline}
with the self energy
\be
\Sigma_b^R(\omega) = -\frac{y^2}{2}\frac{\delta_a}{(\omega +
  i\kappa)^2 - \delta_a^2}\,.
\ee
The advanced Green's function is calculated from
the symmetry ${\bf G}^A_{b,2\times{}2} = [{\bf G}^R_{b,2\times{}2}]^\dagger$, 
and ${\bf D}^K_{2b,\times{}2}$ is given by Eq.~(\ref{eq:D_Kb}).

The interesting part of the Keldysh Green's function is its first
diagonal element which gives the Fourier transform of the atom
correlation function $C_b(t) = \langle \{b(t), b^\dagger(0)\}\rangle$,
that is $C_b(\omega) = [{\bf G}^K_{b,2\times{}2}]_{11}$, 
\begin{multline}
\label{eq:correl_func_b}
C_b(\omega) = \frac{-i}{|{\rm det}[{\bf G}^R_{b,2\times{}2}]^{-1}|^2} \Big( D(\omega) \big[ |[G_b^R]^{-1}(-\omega)|^2 \\
+2{\rm Re}\left\{\Sigma_b^R(\omega)[G_b^R]^{-1}(-\omega)\right\} +
  |\Sigma^R(\omega)|^2\big] \\ 
  D(-\omega)|\Sigma^R(\omega)|^2 + d(\omega)|[G_b^R]^{-1}(-\omega)|^2 \Big)\,.
\end{multline}

The top panel of Fig.~\ref{fig:correl_func_b} explores how the
correlation function of mode $b$ changes when $y$ approaches $y_c$.
Upon the color map we plot the real part of the soft mode pole (shown
in Fig.~\ref{fig:softmode}). Although the pole gives the position of
the resonance quite accurately, we stress that these peaks are not
Lorenzian due to the frequency-dependent numerator of the correlation
function, Eq.~(\ref{eq:correl_func_b}). As increasing the coupling,
the single non-Lorentzian peak for $y=0$ moves towards zero, and a
second peak appears on the negative frequency side due to the
interaction with the cavity photons. Approaching the critical point,
the two peaks merge to form a diverging peak at zero frequency. In the
bottom panel of Fig.~\ref{fig:correl_func_b}, we plot two sets of
correlation functions, for $s=0.8$ (thick lines) and $s=1.8$ (thin
lines), in order to demonstrate the quasi-periodicity of the
correlation function with respect to the bath exponent, explained
earlier in connection with Eq.~(\ref{eq:rho}). The trigonometric
factor in the expression (\ref{eq:K_R_A}) of $K^R(\omega>0)$ is
periodic, and remains the same for $s \rightarrow s+1$, only the
exponent of $\omega$ is different in the two cases. We observe that
the central peak is sharper for the smaller $s$ value. The correlation
functions give the same value at $\omega=0$ for the two different
exponents $s$, since the mode density of the colored bath vanishes
anyway at zero frequency.

\subsection{Critical exponent}

As we indicated before, the critical exponent can be calculated only directly from the diverging fluctuations by determining the power law exponent of the divergence. The excitation numbers of mode $a$ and $b$ in the steady state are the equal-time correlations, i.e. $\langle a^\dagger(0) a(0)
\rangle = (C_a(t=0) - 1)/2$, which can be calculated by integrating the correlation functions in Fourier space
\begin{equation}
\label{eq:phnum}
C_{a,b}(t=0) = \int \frac{\ud\omega}{2\pi} C_{a,b}(\omega)\,.
\end{equation}

\begin{figure}[t!]
\includegraphics[angle=0,width=0.75\columnwidth]{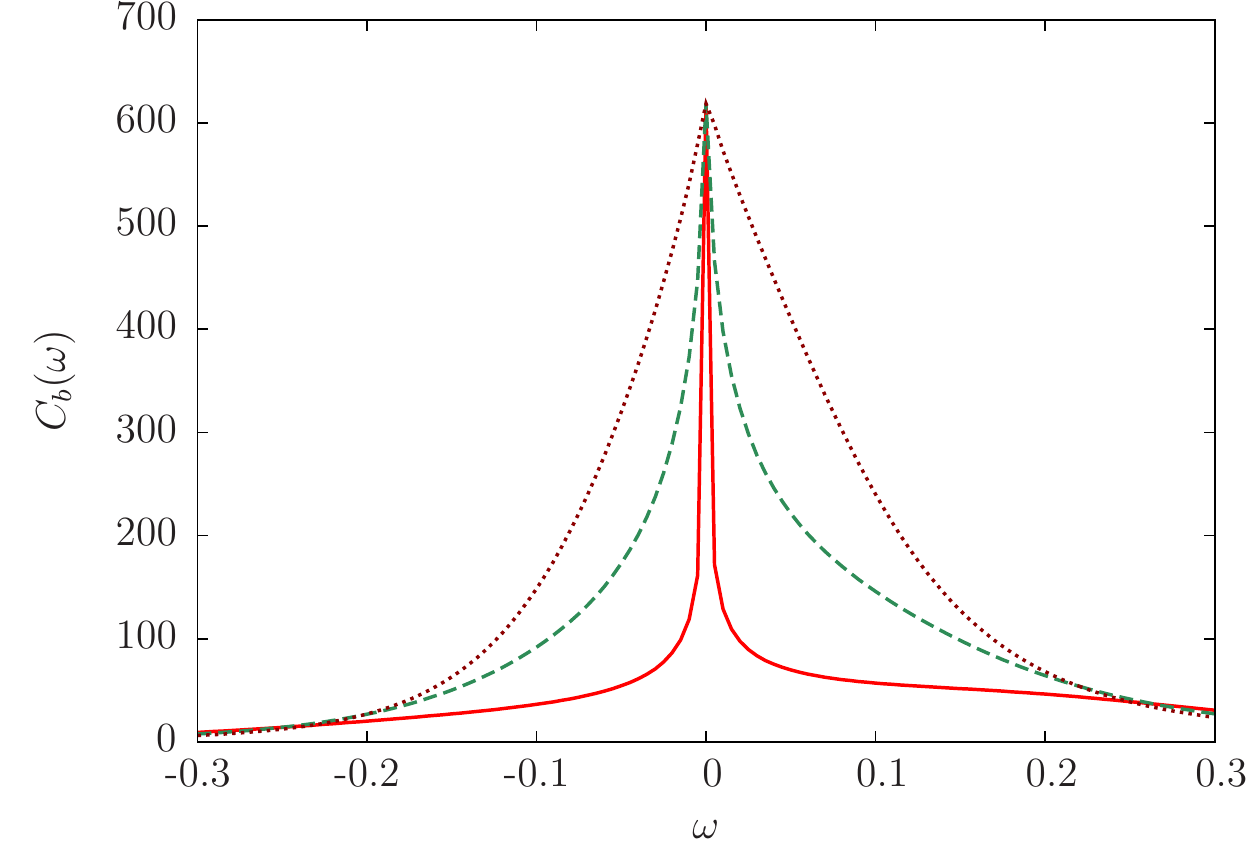}
\caption{(Color online) Atomic correlation functions close to the
  critical point for $y = 0.99 y_c$.  The width of the peak decreases
  with the parameter $s$, that is $s=0.7$ (dotted brown), $s=0.5$
  (dashed green) and $s=0.3$ (solid red). Other parameters are the
  same as in Fig.~\ref{fig:critical_exponent}.}
\label{fig:peaks}
\end{figure}
Let us investigate the atomic correlation function $C_b(\omega)$ at a coupling very close to the critical point ($y=0.99y_c$), which is shown in Figure~\ref{fig:peaks} for three different values of the exponent $s$. These spectra are manifestly far from a Lorentzian form, indicating that the soft mode picture cannot account for the divergence of the fluctuations. The peak points match at zero frequency, at the same time, the smaller the exponent $s$, the narrower the peak. Hence, the area below the peaks, giving the integral in Eq.~(\ref{eq:phnum}), varies as a function of the exponent $s$.

\begin{figure}[t!]
\includegraphics[angle=0,width=0.75\columnwidth]{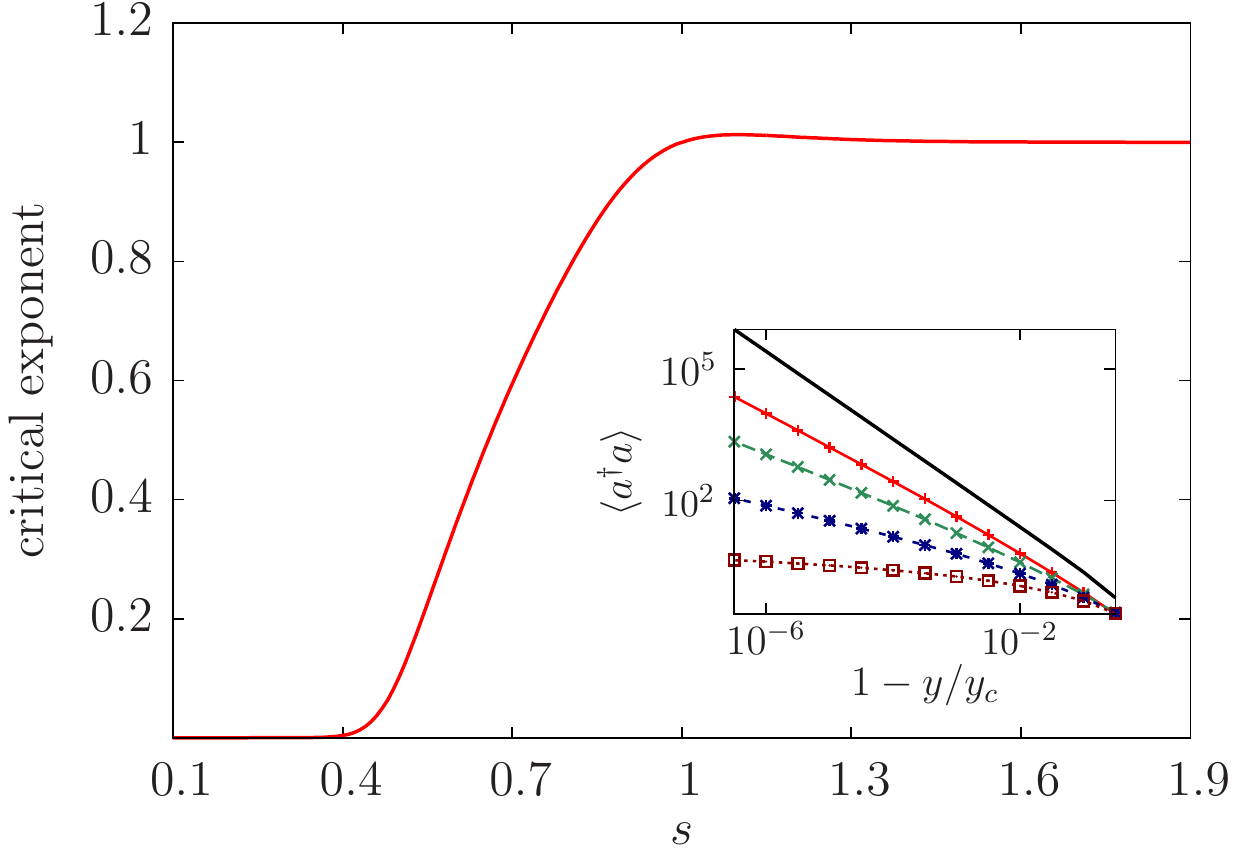}
\caption{(Color online)  Critical exponent as a function of the bath exponent $s$ of
  the colored reservoir. The inset shows the power-law scaling of the
  photon number. Parameters: $\delta_a=2\omega_b$, $\kappa=0.5\omega_b$,
  $\gamma=0.1\omega_b$, $\mu=0$.}
\label{fig:critical_exponent}
\end{figure}
The exponent is extracted from a power-law fit on the excitation
numbers plotted against $1-y/y_c$ close to the critical point $|y -
y_c| < 10^{-4}$. In the inset of Fig.~\ref{fig:critical_exponent}, we
show the scaling of the photon number for  a set of  values of the
parameter $s$.  Even in the presence of the colored noise, one clearly
recognizes the power-law divergence in the close vicinity of the
critical point.

The main panel of Fig.~\ref{fig:critical_exponent} presents the
central result of our paper, where we explore how the critical
exponent of the system varies as a function of the bath exponent
$s$. For $s>1$, we recover the exponent $1$, that is the original
open-system exponent of the Dicke model \cite{Nagy2011Critical}. This
thermal-like exponent arises due to the Markovian dissipation process,
where the fluctuations are Gaussian, similarly to thermal
fluctuations. We find that a super-Ohmic reservoir does not affect
this critical exponent.  In contrast, the critical fluctuations are
significantly reshaped in the sub-Ohmic case for $s<1$, where we find
a monotonic decrease of the critical exponent with decreasing bath parameter
$s$. Moreover, for $s<0.4$, the critical fluctuations are completely
suppressed by the colored reservoir, and $\langle a^\dagger a\rangle$
and $\langle b^\dagger b\rangle$ remain finite at the critical
point. This means that the driving and the photon loss induced noise of the cavity mode is
effectively dissipated into the colored reservoir.

\section{Thermal effects}

There has been a discussion whether the phase transition can be
considered a classical thermal one in the open-system Dicke
model. Although the cavity fluctuations are Gaussian, similarly to
thermal fluctuations,  certain properties of the critical
behaviour make this identification doubtful. In particular, the cavity
and atomic subsystems have different effective temperatures in the
steady-state, hence the driven system \emph{as a whole} cannot be
considered to be in a thermal equilibrium \cite{Torre2013Keldysh}.

Within the Keldysh formalism, we can consider the thermal effects when the low-frequency colored bath has
a finite temperature. For instance, in the experimental realization,
the thermal fraction of the BEC corresponds to the thermal occupation
of the phonon modes that form the colored reservoir. In our general
theory, we include the temperature in the inverse Keldysh component of
mode $b$, Eq.~(\ref{eq:G_K_b}). Meanwhile, we keep the Green's
functions of mode $a$ unchanged, because the temperature is negligible
at the pump frequency $\hbar\omega_p \gg k_B T$. It is only relevant
for the critical dynamics at low frequencies.

\subsection{Thermal equilibrium of mode $b$}

Before turning to the coupled system, we investigate the thermal
average of the population in mode $b$ coupled to a finite-temperature colored
bath. The temperature of the reservoir is included via the Keldysh
component Eq.~(\ref{eq:G_K_b}) in the distribution function $F(\omega)
= \coth\{\hbar\omega/k_BT\}$. The thermal occupation of oscillator $b$
is calculated as previously described. When simplifying the
correlation function Eq.~(\ref{eq:correl_func_b}) for $y=0$, we
obtain the following integral for the zero-time correlation function
(according to Eq.~({\ref{eq:phnum}})), \be
\label{eq:b_thermal}
C_b(t=0) = \int d\omega
\frac{\rho(\omega) \coth(\hbar\omega/2k_BT)}{(\omega-\omega_b-{\rm
    Re}K(\omega))^2 + ({\rm Im}K(\omega))^2}\,.  
\ee 
In order to get the usual thermal occupation number of the oscillator,
one has to assume that the denominator of the integrand has a sharp
peak, so that the cotangent hyperbolic function can be taken out of
the integral. Accordingly, it should be a flat function around the
resonance $\omega_r$, given by $\omega_r-\omega_b-{\rm
  Re}K(\omega_r) = 0$, that leads to the condition
$|\gamma_r\coth^{\prime}(\hbar\omega_r/2k_BT))|\ll
\coth(\hbar\omega_r/2k_BT))$ with $\gamma_r = {\rm
  Im}K(\omega_r)$. Also, the coupling-density profile has to be a flat
function of $\omega$ at resonance,
$|\gamma_r\rho^{\prime}(\omega_r)|\ll \rho(\omega_r)$. With these
approximations one obtains the expected thermal occupation of the
oscillator
\be
\label{eq:phnum_eq}
C_b(t=0) \xrightarrow{\gamma_r\rightarrow 0}
\coth(\hbar\tilde{\omega}_b/2k_BT)\,,
\ee 
with the renormalized oscillator frequency given by the resonance
$\tilde{\omega}_b=\omega_r$. 
The limit $\gamma_r\rightarrow 0$ ensures that the remaining integral 
\be 
\int d\omega \frac{\rho(\omega)}{(\omega-\omega_b-{\rm
    Re}K(\omega))^2 + ({\rm Im}K(\omega))^2} \longrightarrow 1\,.
\ee

\begin{figure}[t!]
\includegraphics[angle=0,width=0.75\columnwidth]{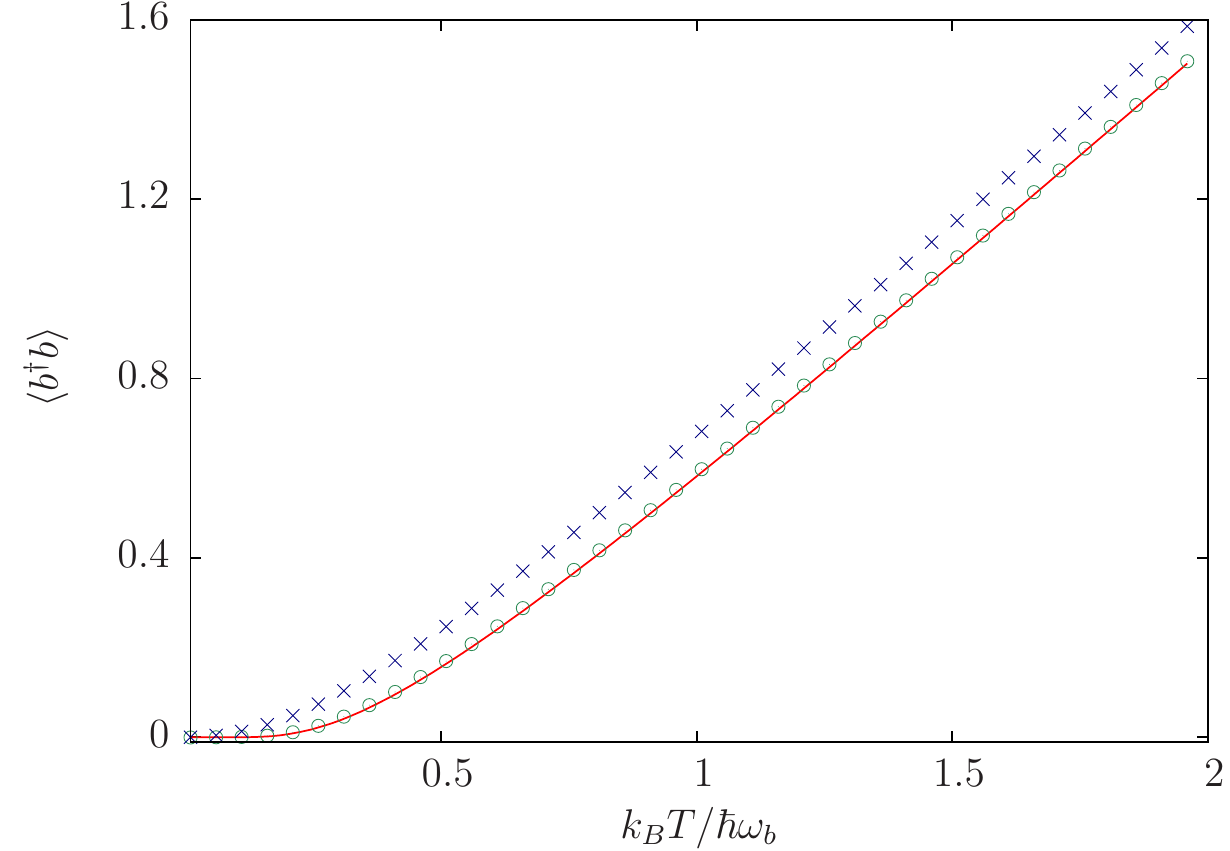}
\caption{(Color online) Thermal occupation of mode $b$ calculated from the
  correlation function at $y=0$ as a function of the bath
  temperature. The analytical curve depicts the thermal state (solid
  red line), which is reached for small reservoir coupling
  $\gamma=0.03\omega_b$ (green circles). For $\gamma=0.5\omega_b$
  (blue crosses), higher occupation numbers are obtained for the colored reservoir. Parameters: 
  $s = 0.6$ and $\mu=0$.}
\label{fig:thermalization}
\end{figure}

In Fig.~\ref{fig:thermalization}, we compare the excitation numbers
in mode $b$ calculated numerically from the integral
Eq.~(\ref{eq:b_thermal}) to the thermal equilibrium value,
Eq.~(\ref{eq:phnum_eq}) without the frequency renormalisation, i.e.,  $\tilde\omega_b \simeq \omega_b$ (solid red line).  When the
coupling to the reservoir is large $\gamma = 0.5 \omega_b$ (blue
crosses) the excitation number is noticabley above the thermal occupation, while
for small reservoir coupling ($\gamma = 0.03\omega_b$, green circles)
the expected  analytical  curve of thermal equilibrium is well reproduced.

Note that, although the Keldysh path integral approach is not
perturbative in $\gamma$ when we integrate out the reservoir modes, in
the end, we still need to consider small coupling to the reservoir so
as to fulfill the thermodynamics of the oscillator. This is related to
the initial assumption that the bath is in thermal equilibrium. Once the system is strongly coupled to the reservoir, the system back-action significantly changes the dynamics of the reservoir modes and the thermal
equilibrium is lost. Normally, realistic decay rates are orders of
magnitude smaller than the oscillator frequency ($\gamma \ll
\omega_b$). The numerical calculations, however, are easier to carry out
for larger decay rates, and this will be our choice.

\subsection{Thermal cross-excitation effects}

\begin{figure}[t!]
\includegraphics[angle=0,width=0.75\columnwidth]{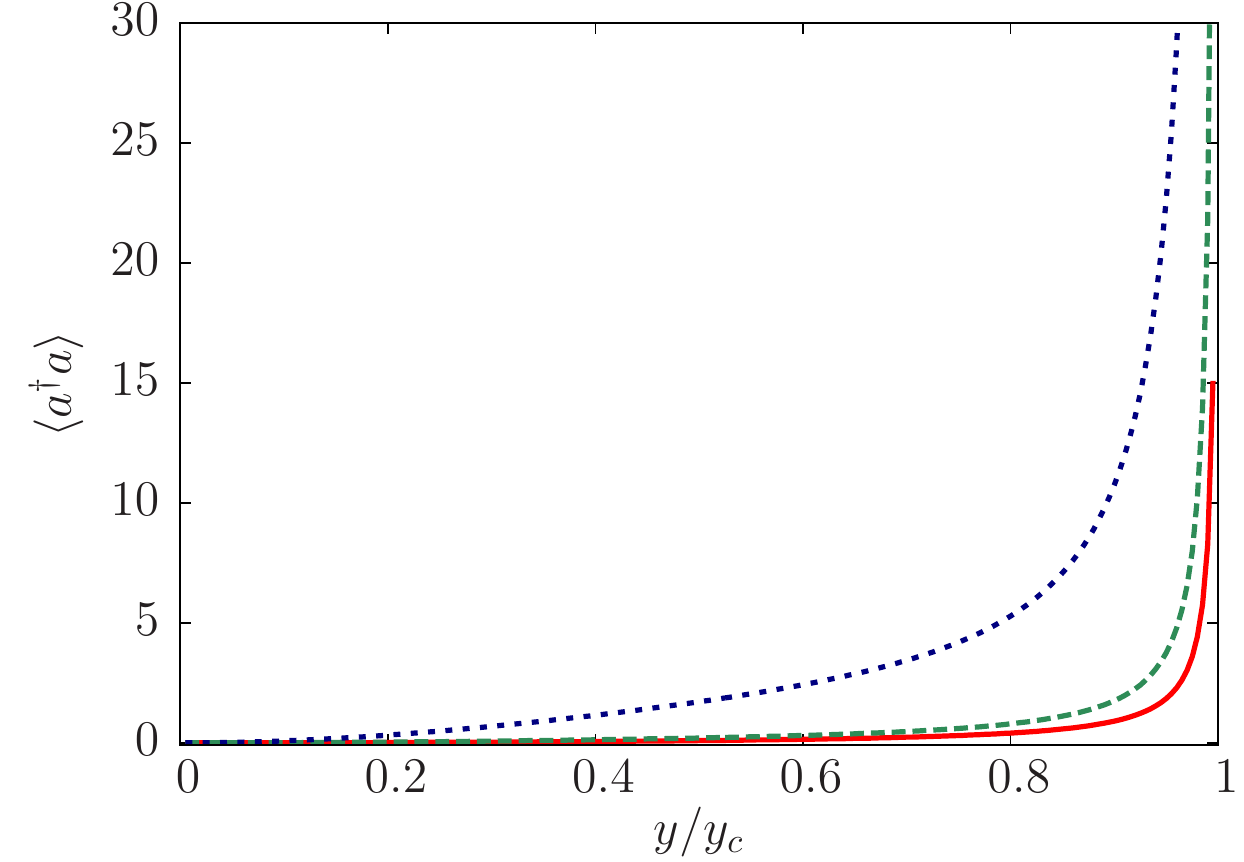}
\includegraphics[angle=0,width=0.75\columnwidth]{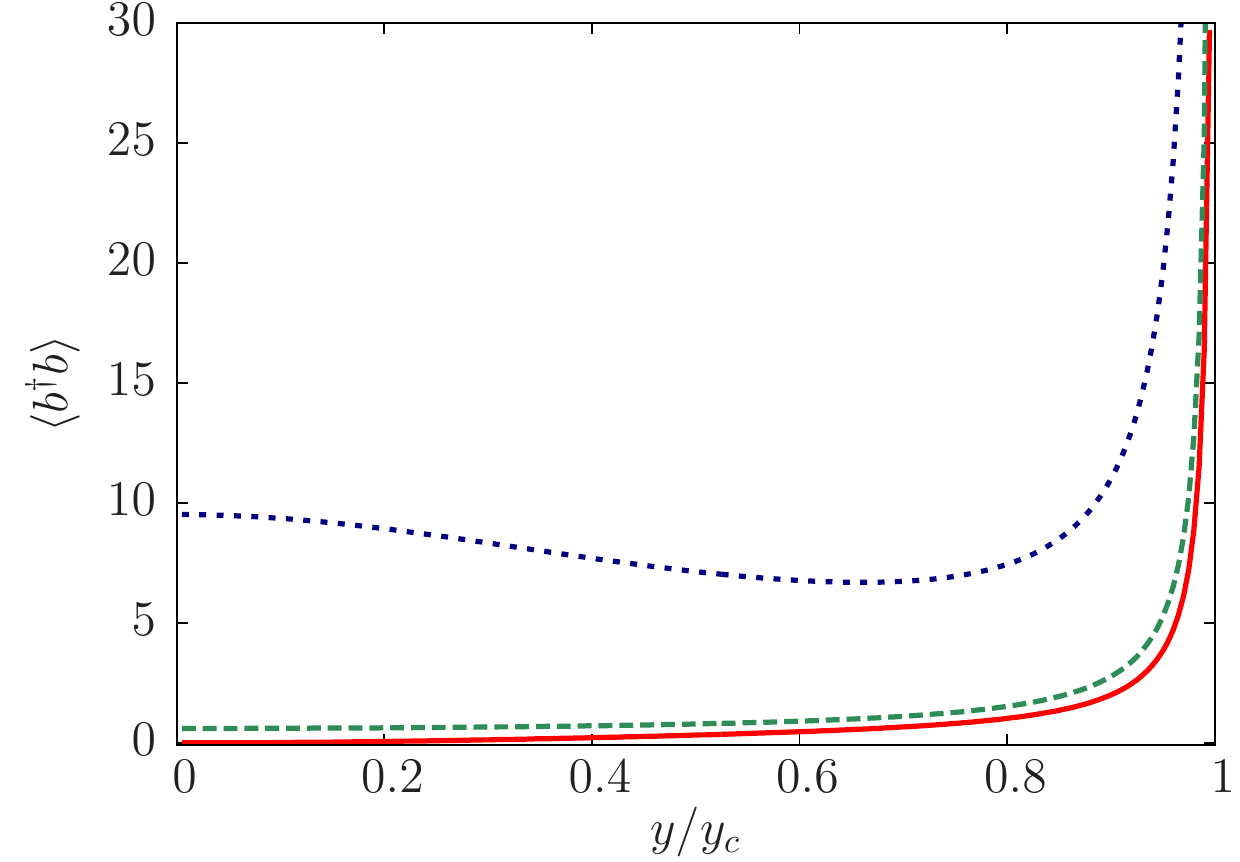}
\caption{(Color online) Excitation number in the cavity mode $a$ (top panel) and the
  atomic mode $b$ (bottom panel) as a function of the relative
  coupling strength $y/y_c$ at different temperatures $T = 0$, $1$,
  and $10$ $\hbar\omega_b/k_B$. Parameters: $\delta_a = 2$,
  $\kappa=2$, $g=0.1$ $\omega_b$.}
\label{fig:thermal_vs_y}
\end{figure}

In Figure~\ref{fig:thermal_vs_y} the cavity photon number is plotted as a function of the relative coupling $y/y_c$ for different temperatures of the colored bath (top panel).
The thermal excitation of mode $b$ gives rise to an increase of the
cavity photon number through photon scattering. This cross-effect reflects again the driving as a substantial feature of the system.  High-frequency photons can be created from the low-frequency thermal excitations. The photon number curve is pushed upward from the $T=0$ curve (solid red); the larger the $T$, the larger the shift. 

Similar behaviour can be seen on the atomic excitation number (bottom panel) for low $T<1$.  The interesting new feature takes place at higher temperature, $k_B T=10 \hbar \omega_b$: the population in mode $b$
decreases with the atom-cavity coupling before the diverging peak at
the critical point. This effect can be attributed to the cavity
cooling mechanism. The cavity dissipation channel through the atom-photon coupling can extract energy
from the atomic mode. Then, the new excitation number is the result of the
competition between the cavity cooling and thermalization. Here we
stress that the theory is valid for small system-bath coupling, which
ensures that the bath can remain in thermal equilibrium.

\subsection{Critical exponent}

\begin{figure}[t!]
  \includegraphics[angle=0,width=0.75\columnwidth]{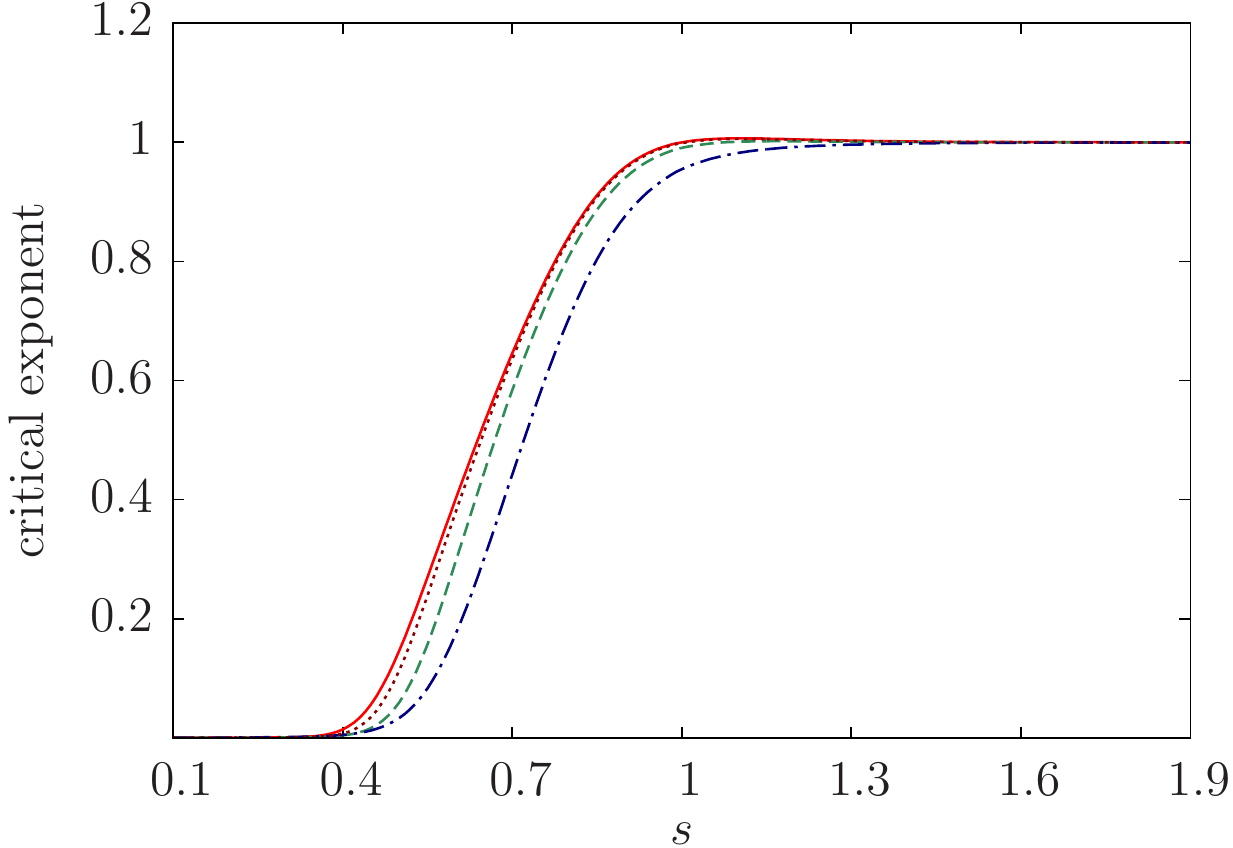}
  \caption{(Color online) Critical exponent as a function of the bath exponent $s$ of
    the colored reservoir for different chemical potentials $\mu =
    -0.1$ (dotted brown), $-0.01$ (dashed green) and $-0.001\omega_b$
    (dashed-dotted blue). Other parameters are $\kappa = 0.5\omega_b$,
    $\gamma = 0.1 \omega_b$, $T = 1\hbar\omega_b/k_B$. We plot the
    zero temperature curve (solid red) of
    Fig.~\ref{fig:critical_exponent} for reference.}
  \label{fig:critical_exponent_finiteT}
\end{figure}

Finally, we study the role of temperature on the criticality of the
driven-damped system. We wish to explore how the critical exponent changes with the bath exponent 
when the colored reservoir has a finite temperature $T = \hbar\omega_b/k_B$. 
In this analysis, the chemical potential of the colored reservoir, introduced in Eq.~(\ref{eq:G_K_b}), plays a significant role below $s=1.3$. Deeply in the super-Ohmic regime $s>1.3$, the same zero-temperature exponent $1$ is obtained for $\mu=0$. For $s<1.3$, our calculation
breaks down indicating that the Fourier transform of the correlation
function is divergent. Finite chemical potential has to be introduced in order to account for the state  of the low-frequency reservoir modes, i.e., in order to  regularize the
diverging particle number in the reservoir modes around $\omega=0$. We performed the calculation on the diverging population in the modes at several chemical potential choices.  Figure~\ref{fig:critical_exponent_finiteT}  shows that the overall behaviour of the critical exponent as a function of the exponent of the reservoir spectral density function is the same as at $T=0$. The overall behaviour of the plotted curves is also unaffected by the chemical potential. For larger chemical potential (dotted brown), the zero-temperature result (solid red) is obtained.  For smaller chemical potential, the central part of the curve is shifted a bit more towards lower
critical exponents, all this happening in the interval $0.4 < s
< 1.3$. In the super-Ohmic regime (above $s>1.3$), the cavity dissipation dominates the critical exponent, and the colored bath proves to be irrelevant regarding the criticality. 

\section{Conclusion and Outlook}

In this paper we studied the effects of a colored reservoir on the criticality occurring in  a 
driven-dissipative system. Our model corresponds to the
cavity QED realization of the Dicke model, where a high-frequency
driven cavity mode is coupled to a low-frequency condensate excitation
mode. We demonstrated that the critical exponent of the superradiant
phase transition is highly sensitive to the spectral features of the
low-frequency reservoir. In particular, we found that the critical
exponent varies monotonously as a function of the bath exponent in the domain of 
sub-Ohmic reservoirs. The criticality vanishes completely at a given value of the bath exponent. For super-Ohmic reservoirs, the critical exponent is determined by the Markovian loss channel and is constant 1. We showed that the non-vanishing temperature does not break this dependence on the spectral density function of the bath, and non-trivial exponents are obtained for sub-Ohmic reservoirs.

Besides the field of cavity QED there are other systems whose
driven-dissipative nature is inherent to their physical
realization. For instance, Dicke-type systems can be designed in
circuit QED
\cite{Nataf2012Double,Baksic2014Controlling,Lolli2015Ancillary}, where
non-Gaussian noise appears from the motion of electrons
\cite{Paladino20141f}. Other examples for out-of-equilibrium systems
are polariton Bose-Einstein condensates created in semiconductor
heterostructures, where dissipation is intrinsic to the dynamics
\cite{Carusotto2013Quantum,Wachtel2016Electrodynamic}.

\begin{acknowledgments}
The authors thank fruitful discussions with Z. Kurucz and G. K\'onya.
This work was supported by the Hungarian Academy of Sciences
(Lend\"ulet Program, LP2011-016) and the National Research,
Development and Innovation Office (K115624). D. N. acknowledges
support from the J\'{a}nos Bolyai Research Scholarship of the
Hungarian Academy of Sciences.
\end{acknowledgments}


\begin{thebibliography}{51}
\expandafter\ifx\csname natexlab\endcsname\relax\def\natexlab#1{#1}\fi
\expandafter\ifx\csname bibnamefont\endcsname\relax
  \def\bibnamefont#1{#1}\fi
\expandafter\ifx\csname bibfnamefont\endcsname\relax
  \def\bibfnamefont#1{#1}\fi
\expandafter\ifx\csname citenamefont\endcsname\relax
  \def\citenamefont#1{#1}\fi
\expandafter\ifx\csname url\endcsname\relax
  \def\url#1{\texttt{#1}}\fi
\expandafter\ifx\csname urlprefix\endcsname\relax\def\urlprefix{URL }\fi
\providecommand{\bibinfo}[2]{#2}
\providecommand{\eprint}[2][]{\url{#2}}

\bibitem[{\citenamefont{Dalla~Torre et~al.}(2010)\citenamefont{Dalla~Torre,
  Demler, Giamarchi, and Altman}}]{DallaTorre2010Quantum}
\bibinfo{author}{\bibfnamefont{E.~G.} \bibnamefont{Dalla~Torre}},
  \bibinfo{author}{\bibfnamefont{E.}~\bibnamefont{Demler}},
  \bibinfo{author}{\bibfnamefont{T.}~\bibnamefont{Giamarchi}},
  \bibnamefont{and} \bibinfo{author}{\bibfnamefont{E.}~\bibnamefont{Altman}},
  \bibinfo{journal}{Nature Physics} \textbf{\bibinfo{volume}{6}},
  \bibinfo{pages}{806} (\bibinfo{year}{2010}).

\bibitem[{\citenamefont{Diehl et~al.}(2010)\citenamefont{Diehl, Tomadin,
  Micheli, Fazio, and Zoller}}]{Diehl2010Dynamical}
\bibinfo{author}{\bibfnamefont{S.}~\bibnamefont{Diehl}},
  \bibinfo{author}{\bibfnamefont{A.}~\bibnamefont{Tomadin}},
  \bibinfo{author}{\bibfnamefont{A.}~\bibnamefont{Micheli}},
  \bibinfo{author}{\bibfnamefont{R.}~\bibnamefont{Fazio}}, \bibnamefont{and}
  \bibinfo{author}{\bibfnamefont{P.}~\bibnamefont{Zoller}},
  \bibinfo{journal}{Phys. Rev. Lett.} \textbf{\bibinfo{volume}{105}},
  \bibinfo{pages}{015702} (\bibinfo{year}{2010}).

\bibitem[{\citenamefont{Dalla~Torre et~al.}(2012)\citenamefont{Dalla~Torre,
  Demler, Giamarchi, and Altman}}]{DallaTorre2012Dynamics}
\bibinfo{author}{\bibfnamefont{E.~G.} \bibnamefont{Dalla~Torre}},
  \bibinfo{author}{\bibfnamefont{E.}~\bibnamefont{Demler}},
  \bibinfo{author}{\bibfnamefont{T.}~\bibnamefont{Giamarchi}},
  \bibnamefont{and} \bibinfo{author}{\bibfnamefont{E.}~\bibnamefont{Altman}},
  \bibinfo{journal}{Phys. Rev. B} \textbf{\bibinfo{volume}{85}},
  \bibinfo{pages}{184302} (\bibinfo{year}{2012}).

\bibitem[{\citenamefont{Chitra and Zilberberg}(2015)}]{Chitra2015Dynamical}
\bibinfo{author}{\bibfnamefont{R.}~\bibnamefont{Chitra}} \bibnamefont{and}
  \bibinfo{author}{\bibfnamefont{O.}~\bibnamefont{Zilberberg}},
  \bibinfo{journal}{Phys. Rev. A} \textbf{\bibinfo{volume}{92}},
  \bibinfo{pages}{023815} (\bibinfo{year}{2015}).

\bibitem[{\citenamefont{Schir\'{o} et~al.}(2016)\citenamefont{Schir\'{o},
  Joshi, Bordyuh, Fazio, Keeling, and T\"{u}reci}}]{Schiro2016Exotic}
\bibinfo{author}{\bibfnamefont{M.}~\bibnamefont{Schir\'{o}}},
  \bibinfo{author}{\bibfnamefont{C.}~\bibnamefont{Joshi}},
  \bibinfo{author}{\bibfnamefont{M.}~\bibnamefont{Bordyuh}},
  \bibinfo{author}{\bibfnamefont{R.}~\bibnamefont{Fazio}},
  \bibinfo{author}{\bibfnamefont{J.}~\bibnamefont{Keeling}}, \bibnamefont{and}
  \bibinfo{author}{\bibfnamefont{H.~E.} \bibnamefont{T\"{u}reci}},
  \bibinfo{journal}{Phys. Rev. Lett.} \textbf{\bibinfo{volume}{116}},
  \bibinfo{pages}{143603} (\bibinfo{year}{2016}).

\bibitem[{\citenamefont{Sieberer et~al.}(2013)\citenamefont{Sieberer, Huber,
  Altman, and Diehl}}]{Sieberer2013Dynamical}
\bibinfo{author}{\bibfnamefont{L.~M.} \bibnamefont{Sieberer}},
  \bibinfo{author}{\bibfnamefont{S.~D.} \bibnamefont{Huber}},
  \bibinfo{author}{\bibfnamefont{E.}~\bibnamefont{Altman}}, \bibnamefont{and}
  \bibinfo{author}{\bibfnamefont{S.}~\bibnamefont{Diehl}},
  \bibinfo{journal}{Phys. Rev. Lett.} \textbf{\bibinfo{volume}{110}},
  \bibinfo{pages}{195301} (\bibinfo{year}{2013}).

\bibitem[{\citenamefont{Marino and
  Diehl}(2016{\natexlab{a}})}]{Marino2016Driven}
\bibinfo{author}{\bibfnamefont{J.}~\bibnamefont{Marino}} \bibnamefont{and}
  \bibinfo{author}{\bibfnamefont{S.}~\bibnamefont{Diehl}},
  \bibinfo{journal}{Phys. Rev. Lett.} \textbf{\bibinfo{volume}{116}},
  \bibinfo{pages}{070407} (\bibinfo{year}{2016}{\natexlab{a}}).

\bibitem[{\citenamefont{Marino and
  Diehl}(2016{\natexlab{b}})}]{Marino2016Quantum}
\bibinfo{author}{\bibfnamefont{J.}~\bibnamefont{Marino}} \bibnamefont{and}
  \bibinfo{author}{\bibfnamefont{S.}~\bibnamefont{Diehl}},
  \bibinfo{journal}{Phys. Rev. B} \textbf{\bibinfo{volume}{94}},
  \bibinfo{pages}{085150} (\bibinfo{year}{2016}{\natexlab{b}}).

\bibitem[{\citenamefont{Lang and Piazza}(2016)}]{Lang2016Critical}
\bibinfo{author}{\bibfnamefont{J.}~\bibnamefont{Lang}} \bibnamefont{and}
  \bibinfo{author}{\bibfnamefont{F.}~\bibnamefont{Piazza}},
  \bibinfo{journal}{Phys. Rev. A} \textbf{\bibinfo{volume}{94}},
  \bibinfo{pages}{033628} (\bibinfo{year}{2016}).

\bibitem[{\citenamefont{Le~Boit\'{e} et~al.}(2013)\citenamefont{Le~Boit\'{e},
  Orso, and Ciuti}}]{LeBoite2013SteadyState}
\bibinfo{author}{\bibfnamefont{A.}~\bibnamefont{Le~Boit\'{e}}},
  \bibinfo{author}{\bibfnamefont{G.}~\bibnamefont{Orso}}, \bibnamefont{and}
  \bibinfo{author}{\bibfnamefont{C.}~\bibnamefont{Ciuti}},
  \bibinfo{journal}{Phys. Rev. Lett.} \textbf{\bibinfo{volume}{110}},
  \bibinfo{pages}{233601} (\bibinfo{year}{2013}).

\bibitem[{\citenamefont{Le~Boit\'{e} et~al.}(2014)\citenamefont{Le~Boit\'{e},
  Orso, and Ciuti}}]{LeBoite2014BoseHubbard}
\bibinfo{author}{\bibfnamefont{A.}~\bibnamefont{Le~Boit\'{e}}},
  \bibinfo{author}{\bibfnamefont{G.}~\bibnamefont{Orso}}, \bibnamefont{and}
  \bibinfo{author}{\bibfnamefont{C.}~\bibnamefont{Ciuti}},
  \bibinfo{journal}{Phys. Rev. A} \textbf{\bibinfo{volume}{90}},
  \bibinfo{pages}{063821} (\bibinfo{year}{2014}).

\bibitem[{\citenamefont{Griess{}er and
  Ritsch}(2013)}]{Griesser2013Lightinduced}
\bibinfo{author}{\bibfnamefont{T.}~\bibnamefont{Griess{}er}} \bibnamefont{and}
  \bibinfo{author}{\bibfnamefont{H.}~\bibnamefont{Ritsch}},
  \bibinfo{journal}{Phys. Rev. Lett.} \textbf{\bibinfo{volume}{111}},
  \bibinfo{pages}{055702} (\bibinfo{year}{2013}).

\bibitem[{\citenamefont{Ostermann et~al.}(2016)\citenamefont{Ostermann, Piazza,
  and Ritsch}}]{Ostermann2016Spontaneous}
\bibinfo{author}{\bibfnamefont{S.}~\bibnamefont{Ostermann}},
  \bibinfo{author}{\bibfnamefont{F.}~\bibnamefont{Piazza}}, \bibnamefont{and}
  \bibinfo{author}{\bibfnamefont{H.}~\bibnamefont{Ritsch}},
  \bibinfo{journal}{Phys. Rev. X} \textbf{\bibinfo{volume}{6}},
  \bibinfo{pages}{021026} (\bibinfo{year}{2016}).

\bibitem[{\citenamefont{Sachdev}(2011)}]{Sachdev2011Quantum}
\bibinfo{author}{\bibfnamefont{S.}~\bibnamefont{Sachdev}},
  \emph{\bibinfo{title}{{Quantum Phase Transitions}}}
  (\bibinfo{publisher}{Cambridge University Press}, \bibinfo{year}{2011}), ISBN
  \bibinfo{isbn}{978-0-521-51468-2}.

\bibitem[{\citenamefont{Strack and Sachdev}(2011)}]{Strack2011Dicke}
\bibinfo{author}{\bibfnamefont{P.}~\bibnamefont{Strack}} \bibnamefont{and}
  \bibinfo{author}{\bibfnamefont{S.}~\bibnamefont{Sachdev}},
  \bibinfo{journal}{Phys. Rev. Lett.} \textbf{\bibinfo{volume}{107}},
  \bibinfo{pages}{277202} (\bibinfo{year}{2011}).

\bibitem[{\citenamefont{Piazza and Strack}(2014)}]{Piazza2014Umklapp}
\bibinfo{author}{\bibfnamefont{F.}~\bibnamefont{Piazza}} \bibnamefont{and}
  \bibinfo{author}{\bibfnamefont{P.}~\bibnamefont{Strack}},
  \bibinfo{journal}{Phys. Rev. Lett.} \textbf{\bibinfo{volume}{112}},
  \bibinfo{pages}{143003} (\bibinfo{year}{2014}).

\bibitem[{\citenamefont{Hwang et~al.}(2015)\citenamefont{Hwang, Puebla, and
  Plenio}}]{Hwang2015Quantum}
\bibinfo{author}{\bibfnamefont{M.~J.} \bibnamefont{Hwang}},
  \bibinfo{author}{\bibfnamefont{R.}~\bibnamefont{Puebla}}, \bibnamefont{and}
  \bibinfo{author}{\bibfnamefont{M.~B.} \bibnamefont{Plenio}},
  \bibinfo{journal}{Phys. Rev. Lett.} \textbf{\bibinfo{volume}{115}},
  \bibinfo{pages}{180404} (\bibinfo{year}{2015}).

\bibitem[{\citenamefont{Niederle et~al.}(2016)\citenamefont{Niederle, Morigi,
  and Rieger}}]{Niederle2016Ultracold}
\bibinfo{author}{\bibfnamefont{A.~E.} \bibnamefont{Niederle}},
  \bibinfo{author}{\bibfnamefont{G.}~\bibnamefont{Morigi}}, \bibnamefont{and}
  \bibinfo{author}{\bibfnamefont{H.}~\bibnamefont{Rieger}},
  \bibinfo{journal}{Phys. Rev. A} \textbf{\bibinfo{volume}{94}},
  \bibinfo{pages}{033607} (\bibinfo{year}{2016}).

\bibitem[{\citenamefont{Baumann et~al.}(2010)\citenamefont{Baumann, Guerlin,
  Brennecke, and Esslinger}}]{Baumann2010Dicke}
\bibinfo{author}{\bibfnamefont{K.}~\bibnamefont{Baumann}},
  \bibinfo{author}{\bibfnamefont{C.}~\bibnamefont{Guerlin}},
  \bibinfo{author}{\bibfnamefont{F.}~\bibnamefont{Brennecke}},
  \bibnamefont{and}
  \bibinfo{author}{\bibfnamefont{T.}~\bibnamefont{Esslinger}},
  \bibinfo{journal}{Nature} \textbf{\bibinfo{volume}{464}},
  \bibinfo{pages}{1301} (\bibinfo{year}{2010}).

\bibitem[{\citenamefont{Baumann et~al.}(2011)\citenamefont{Baumann, Mottl,
  Brennecke, and Esslinger}}]{Baumann2011Exploring}
\bibinfo{author}{\bibfnamefont{K.}~\bibnamefont{Baumann}},
  \bibinfo{author}{\bibfnamefont{R.}~\bibnamefont{Mottl}},
  \bibinfo{author}{\bibfnamefont{F.}~\bibnamefont{Brennecke}},
  \bibnamefont{and}
  \bibinfo{author}{\bibfnamefont{T.}~\bibnamefont{Esslinger}},
  \bibinfo{journal}{Phys. Rev. Lett.} \textbf{\bibinfo{volume}{107}},
  \bibinfo{pages}{140402} (\bibinfo{year}{2011}).

\bibitem[{\citenamefont{Mottl et~al.}(2012)\citenamefont{Mottl, Brennecke,
  Baumann, Landig, Donner, and Esslinger}}]{Mottl2012RotonType}
\bibinfo{author}{\bibfnamefont{R.}~\bibnamefont{Mottl}},
  \bibinfo{author}{\bibfnamefont{F.}~\bibnamefont{Brennecke}},
  \bibinfo{author}{\bibfnamefont{K.}~\bibnamefont{Baumann}},
  \bibinfo{author}{\bibfnamefont{R.}~\bibnamefont{Landig}},
  \bibinfo{author}{\bibfnamefont{T.}~\bibnamefont{Donner}}, \bibnamefont{and}
  \bibinfo{author}{\bibfnamefont{T.}~\bibnamefont{Esslinger}},
  \bibinfo{journal}{Science} \textbf{\bibinfo{volume}{336}},
  \bibinfo{pages}{1570} (\bibinfo{year}{2012}).

\bibitem[{\citenamefont{Schmidt et~al.}(2014)\citenamefont{Schmidt, Tomczyk,
  Slama, and Zimmermann}}]{Schmidt2014Dynamical}
\bibinfo{author}{\bibfnamefont{D.}~\bibnamefont{Schmidt}},
  \bibinfo{author}{\bibfnamefont{H.}~\bibnamefont{Tomczyk}},
  \bibinfo{author}{\bibfnamefont{S.}~\bibnamefont{Slama}}, \bibnamefont{and}
  \bibinfo{author}{\bibfnamefont{C.}~\bibnamefont{Zimmermann}},
  \bibinfo{journal}{Phys. Rev. Lett.} \textbf{\bibinfo{volume}{112}},
  \bibinfo{pages}{115302} (\bibinfo{year}{2014}).

\bibitem[{\citenamefont{Baden et~al.}(2014)\citenamefont{Baden, Arnold,
  Grimsmo, Parkins, and Barrett}}]{Baden2014Realization}
\bibinfo{author}{\bibfnamefont{M.~P.} \bibnamefont{Baden}},
  \bibinfo{author}{\bibfnamefont{K.~J.} \bibnamefont{Arnold}},
  \bibinfo{author}{\bibfnamefont{A.~L.} \bibnamefont{Grimsmo}},
  \bibinfo{author}{\bibfnamefont{S.}~\bibnamefont{Parkins}}, \bibnamefont{and}
  \bibinfo{author}{\bibfnamefont{M.~D.} \bibnamefont{Barrett}},
  \bibinfo{journal}{Phys. Rev. Lett.} \textbf{\bibinfo{volume}{113}},
  \bibinfo{pages}{020408} (\bibinfo{year}{2014}).

\bibitem[{\citenamefont{Klinder
  et~al.}(2015{\natexlab{a}})\citenamefont{Klinder, Ke{\ss}ler, Wolke, Mathey,
  and Hemmerich}}]{Klinder2015Dynamical}
\bibinfo{author}{\bibfnamefont{J.}~\bibnamefont{Klinder}},
  \bibinfo{author}{\bibfnamefont{H.}~\bibnamefont{Ke{\ss}ler}},
  \bibinfo{author}{\bibfnamefont{M.}~\bibnamefont{Wolke}},
  \bibinfo{author}{\bibfnamefont{L.}~\bibnamefont{Mathey}}, \bibnamefont{and}
  \bibinfo{author}{\bibfnamefont{A.}~\bibnamefont{Hemmerich}},
  \bibinfo{journal}{P. Natl. Acad. Sci. USA} p. \bibinfo{pages}{201417132}
  (\bibinfo{year}{2015}{\natexlab{a}}).

\bibitem[{\citenamefont{Klinder
  et~al.}(2015{\natexlab{b}})\citenamefont{Klinder, Kess{}ler, Bakhtiari,
  Thorwart, and Hemmerich}}]{Klinder2015Observation}
\bibinfo{author}{\bibfnamefont{J.}~\bibnamefont{Klinder}},
  \bibinfo{author}{\bibfnamefont{H.}~\bibnamefont{Kess{}ler}},
  \bibinfo{author}{\bibfnamefont{M.~R.} \bibnamefont{Bakhtiari}},
  \bibinfo{author}{\bibfnamefont{M.}~\bibnamefont{Thorwart}}, \bibnamefont{and}
  \bibinfo{author}{\bibfnamefont{A.}~\bibnamefont{Hemmerich}},
  \bibinfo{journal}{Phys. Rev. Lett.} \textbf{\bibinfo{volume}{115}},
  \bibinfo{pages}{230403} (\bibinfo{year}{2015}{\natexlab{b}}).

\bibitem[{\citenamefont{Landig et~al.}(2016)\citenamefont{Landig, Hruby, Dogra,
  Landini, Mottl, Donner, and Esslinger}}]{Landig2016Quantum}
\bibinfo{author}{\bibfnamefont{R.}~\bibnamefont{Landig}},
  \bibinfo{author}{\bibfnamefont{L.}~\bibnamefont{Hruby}},
  \bibinfo{author}{\bibfnamefont{N.}~\bibnamefont{Dogra}},
  \bibinfo{author}{\bibfnamefont{M.}~\bibnamefont{Landini}},
  \bibinfo{author}{\bibfnamefont{R.}~\bibnamefont{Mottl}},
  \bibinfo{author}{\bibfnamefont{T.}~\bibnamefont{Donner}}, \bibnamefont{and}
  \bibinfo{author}{\bibfnamefont{T.}~\bibnamefont{Esslinger}},
  \bibinfo{journal}{Nature} \textbf{\bibinfo{volume}{532}},
  \bibinfo{pages}{476} (\bibinfo{year}{2016}).

\bibitem[{\citenamefont{Koll\'{a}r et~al.}(2015)\citenamefont{Koll\'{a}r,
  Papageorge, Baumann, Armen, and Lev}}]{Kollar2015Adjustablelength}
\bibinfo{author}{\bibfnamefont{A.~J.} \bibnamefont{Koll\'{a}r}},
  \bibinfo{author}{\bibfnamefont{A.~T.} \bibnamefont{Papageorge}},
  \bibinfo{author}{\bibfnamefont{K.}~\bibnamefont{Baumann}},
  \bibinfo{author}{\bibfnamefont{M.~A.} \bibnamefont{Armen}}, \bibnamefont{and}
  \bibinfo{author}{\bibfnamefont{B.~L.} \bibnamefont{Lev}},
  \bibinfo{journal}{New Journal of Physics} \textbf{\bibinfo{volume}{17}},
  \bibinfo{pages}{043012} (\bibinfo{year}{2015}).

\bibitem[{\citenamefont{Koll\'{a}r et~al.}(2016)\citenamefont{Koll\'{a}r,
  Papageorge, Vaidya, Guo, Keeling, and
  Lev}}]{Kollar2016SupermodeDensityWavePolariton}
\bibinfo{author}{\bibfnamefont{A.~J.} \bibnamefont{Koll\'{a}r}},
  \bibinfo{author}{\bibfnamefont{A.~T.} \bibnamefont{Papageorge}},
  \bibinfo{author}{\bibfnamefont{V.~D.} \bibnamefont{Vaidya}},
  \bibinfo{author}{\bibfnamefont{Y.}~\bibnamefont{Guo}},
  \bibinfo{author}{\bibfnamefont{J.}~\bibnamefont{Keeling}}, \bibnamefont{and}
  \bibinfo{author}{\bibfnamefont{B.~L.} \bibnamefont{Lev}},
  \bibinfo{journal}{arXiv} p. \bibinfo{pages}{1606.04127}
  (\bibinfo{year}{2016}).

\bibitem[{\citenamefont{Brennecke et~al.}(2013)\citenamefont{Brennecke, Mottl,
  Baumann, Landig, Donner, and Esslinger}}]{Brennecke2013Realtime}
\bibinfo{author}{\bibfnamefont{F.}~\bibnamefont{Brennecke}},
  \bibinfo{author}{\bibfnamefont{R.}~\bibnamefont{Mottl}},
  \bibinfo{author}{\bibfnamefont{K.}~\bibnamefont{Baumann}},
  \bibinfo{author}{\bibfnamefont{R.}~\bibnamefont{Landig}},
  \bibinfo{author}{\bibfnamefont{T.}~\bibnamefont{Donner}}, \bibnamefont{and}
  \bibinfo{author}{\bibfnamefont{T.}~\bibnamefont{Esslinger}},
  \bibinfo{journal}{P. Natl. Acad. Sci. USA} \textbf{\bibinfo{volume}{110}},
  \bibinfo{pages}{11763} (\bibinfo{year}{2013}).

\bibitem[{\citenamefont{Vidal and Dusuel}(2007)}]{Vidal2007Finitesize}
\bibinfo{author}{\bibfnamefont{J.}~\bibnamefont{Vidal}} \bibnamefont{and}
  \bibinfo{author}{\bibfnamefont{S.}~\bibnamefont{Dusuel}},
  \bibinfo{journal}{Europhys. Lett.} p. \bibinfo{pages}{817}
  (\bibinfo{year}{2007}).

\bibitem[{\citenamefont{Nagy et~al.}(2010)\citenamefont{Nagy, K\'{o}nya,
  Szirmai, and Domokos}}]{Nagy2010DickeModel}
\bibinfo{author}{\bibfnamefont{D.}~\bibnamefont{Nagy}},
  \bibinfo{author}{\bibfnamefont{G.}~\bibnamefont{K\'{o}nya}},
  \bibinfo{author}{\bibfnamefont{G.}~\bibnamefont{Szirmai}}, \bibnamefont{and}
  \bibinfo{author}{\bibfnamefont{P.}~\bibnamefont{Domokos}},
  \bibinfo{journal}{Phys. Rev. Lett.} \textbf{\bibinfo{volume}{104}},
  \bibinfo{pages}{130401} (\bibinfo{year}{2010}).

\bibitem[{\citenamefont{Nagy et~al.}(2011)\citenamefont{Nagy, Szirmai, and
  Domokos}}]{Nagy2011Critical}
\bibinfo{author}{\bibfnamefont{D.}~\bibnamefont{Nagy}},
  \bibinfo{author}{\bibfnamefont{G.}~\bibnamefont{Szirmai}}, \bibnamefont{and}
  \bibinfo{author}{\bibfnamefont{P.}~\bibnamefont{Domokos}},
  \bibinfo{journal}{Phys. Rev. A} \textbf{\bibinfo{volume}{84}},
  \bibinfo{pages}{043637} (\bibinfo{year}{2011}).

\bibitem[{\citenamefont{Torre et~al.}(2013)\citenamefont{Torre, Diehl, Lukin,
  Sachdev, and Strack}}]{Torre2013Keldysh}
\bibinfo{author}{\bibfnamefont{E.~G.} \bibnamefont{Torre}},
  \bibinfo{author}{\bibfnamefont{S.}~\bibnamefont{Diehl}},
  \bibinfo{author}{\bibfnamefont{M.~D.} \bibnamefont{Lukin}},
  \bibinfo{author}{\bibfnamefont{S.}~\bibnamefont{Sachdev}}, \bibnamefont{and}
  \bibinfo{author}{\bibfnamefont{P.}~\bibnamefont{Strack}},
  \bibinfo{journal}{Phys. Rev. A} \textbf{\bibinfo{volume}{87}},
  \bibinfo{pages}{023831} (\bibinfo{year}{2013}).

\bibitem[{\citenamefont{Nagy and Domokos}(2015)}]{Nagy2015Nonequilibrium}
\bibinfo{author}{\bibfnamefont{D.}~\bibnamefont{Nagy}} \bibnamefont{and}
  \bibinfo{author}{\bibfnamefont{P.}~\bibnamefont{Domokos}},
  \bibinfo{journal}{Phys. Rev. Lett.} \textbf{\bibinfo{volume}{115}},
  \bibinfo{pages}{043601} (\bibinfo{year}{2015}).

\bibitem[{\citenamefont{K\'{o}nya
  et~al.}(2014{\natexlab{a}})\citenamefont{K\'{o}nya, Szirmai, Nagy, and
  Domokos}}]{Konya2014Photonic}
\bibinfo{author}{\bibfnamefont{G.}~\bibnamefont{K\'{o}nya}},
  \bibinfo{author}{\bibfnamefont{G.}~\bibnamefont{Szirmai}},
  \bibinfo{author}{\bibfnamefont{D.}~\bibnamefont{Nagy}}, \bibnamefont{and}
  \bibinfo{author}{\bibfnamefont{P.}~\bibnamefont{Domokos}},
  \bibinfo{journal}{Phys. Rev. A} \textbf{\bibinfo{volume}{89}},
  \bibinfo{pages}{051601} (\bibinfo{year}{2014}{\natexlab{a}}).

\bibitem[{\citenamefont{K\'{o}nya
  et~al.}(2014{\natexlab{b}})\citenamefont{K\'{o}nya, Szirmai, and
  Domokos}}]{Konya2014Damping}
\bibinfo{author}{\bibfnamefont{G.}~\bibnamefont{K\'{o}nya}},
  \bibinfo{author}{\bibfnamefont{G.}~\bibnamefont{Szirmai}}, \bibnamefont{and}
  \bibinfo{author}{\bibfnamefont{P.}~\bibnamefont{Domokos}},
  \bibinfo{journal}{Phys. Rev. A} \textbf{\bibinfo{volume}{90}},
  \bibinfo{pages}{013623} (\bibinfo{year}{2014}{\natexlab{b}}).

\bibitem[{\citenamefont{Ritsch et~al.}(2013)\citenamefont{Ritsch, Domokos,
  Brennecke, and Esslinger}}]{Ritsch2013Cold}
\bibinfo{author}{\bibfnamefont{H.}~\bibnamefont{Ritsch}},
  \bibinfo{author}{\bibfnamefont{P.}~\bibnamefont{Domokos}},
  \bibinfo{author}{\bibfnamefont{F.}~\bibnamefont{Brennecke}},
  \bibnamefont{and}
  \bibinfo{author}{\bibfnamefont{T.}~\bibnamefont{Esslinger}},
  \bibinfo{journal}{Rev. Mod. Phys.} \textbf{\bibinfo{volume}{85}},
  \bibinfo{pages}{553} (\bibinfo{year}{2013}).

\bibitem[{\citenamefont{Garraway}(2011)}]{Garraway2011Dicke}
\bibinfo{author}{\bibfnamefont{B.~M.} \bibnamefont{Garraway}},
  \bibinfo{journal}{Philosophical Transactions of the Royal Society of London
  A: Mathematical, Physical and Engineering Sciences}
  \textbf{\bibinfo{volume}{369}}, \bibinfo{pages}{1137} (\bibinfo{year}{2011}).

\bibitem[{\citenamefont{Cardy}(1996)}]{Cardy1996Scaling}
\bibinfo{author}{\bibfnamefont{J.}~\bibnamefont{Cardy}},
  \emph{\bibinfo{title}{{Scaling and Renormalization in Statistical Physics}}}
  (\bibinfo{publisher}{Cambridge University Press}, \bibinfo{year}{1996}).

\bibitem[{\citenamefont{Emary and Brandes}(2003)}]{Emary2003Chaos}
\bibinfo{author}{\bibfnamefont{C.}~\bibnamefont{Emary}} \bibnamefont{and}
  \bibinfo{author}{\bibfnamefont{T.}~\bibnamefont{Brandes}},
  \bibinfo{journal}{Phys. Rev. E} \textbf{\bibinfo{volume}{67}},
  \bibinfo{pages}{066203} (\bibinfo{year}{2003}).

\bibitem[{\citenamefont{Kurucz and M{\o}{}lmer}(2010)}]{Kurucz2010Multilevel}
\bibinfo{author}{\bibfnamefont{Z.}~\bibnamefont{Kurucz}} \bibnamefont{and}
  \bibinfo{author}{\bibfnamefont{K.}~\bibnamefont{M{\o}{}lmer}},
  \bibinfo{journal}{Phys. Rev. A} \textbf{\bibinfo{volume}{81}},
  \bibinfo{pages}{032314} (\bibinfo{year}{2010}).

\bibitem[{\citenamefont{Cohen-Tannoudji
  et~al.}(1992)\citenamefont{Cohen-Tannoudji, Dupont-Roc, and
  Grinberg}}]{CohenTannoudji1992AtomPhoton}
\bibinfo{author}{\bibfnamefont{C.}~\bibnamefont{Cohen-Tannoudji}},
  \bibinfo{author}{\bibfnamefont{J.}~\bibnamefont{Dupont-Roc}},
  \bibnamefont{and} \bibinfo{author}{\bibfnamefont{G.}~\bibnamefont{Grinberg}},
  \emph{\bibinfo{title}{{Atom--Photon Interactions}}}
  (\bibinfo{publisher}{Wiley}, \bibinfo{address}{New York},
  \bibinfo{year}{1992}).

\bibitem[{\citenamefont{Leggett et~al.}(1987)\citenamefont{Leggett,
  Chakravarty, Dorsey, Fisher, Garg, and Zwerger}}]{Leggett1987Dynamics}
\bibinfo{author}{\bibfnamefont{A.~J.} \bibnamefont{Leggett}},
  \bibinfo{author}{\bibfnamefont{S.}~\bibnamefont{Chakravarty}},
  \bibinfo{author}{\bibfnamefont{A.~T.} \bibnamefont{Dorsey}},
  \bibinfo{author}{\bibfnamefont{M.~P.~A.} \bibnamefont{Fisher}},
  \bibinfo{author}{\bibfnamefont{A.}~\bibnamefont{Garg}}, \bibnamefont{and}
  \bibinfo{author}{\bibfnamefont{W.}~\bibnamefont{Zwerger}},
  \bibinfo{journal}{Rev. Mod. Phys.} \textbf{\bibinfo{volume}{59}},
  \bibinfo{pages}{1} (\bibinfo{year}{1987}).

\bibitem[{\citenamefont{Kessler et~al.}(2012)\citenamefont{Kessler, Giedke,
  Imamoglu, Yelin, Lukin, and Cirac}}]{Kessler2012Dissipative}
\bibinfo{author}{\bibfnamefont{E.~M.} \bibnamefont{Kessler}},
  \bibinfo{author}{\bibfnamefont{G.}~\bibnamefont{Giedke}},
  \bibinfo{author}{\bibfnamefont{A.}~\bibnamefont{Imamoglu}},
  \bibinfo{author}{\bibfnamefont{S.~F.} \bibnamefont{Yelin}},
  \bibinfo{author}{\bibfnamefont{M.~D.} \bibnamefont{Lukin}}, \bibnamefont{and}
  \bibinfo{author}{\bibfnamefont{J.~I.} \bibnamefont{Cirac}},
  \bibinfo{journal}{Phys. Rev. A} \textbf{\bibinfo{volume}{86}},
  \bibinfo{pages}{012116} (\bibinfo{year}{2012}).

\bibitem[{\citenamefont{Eleuch and Rotter}(2013)}]{Eleuch2013Width}
\bibinfo{author}{\bibfnamefont{H.}~\bibnamefont{Eleuch}} \bibnamefont{and}
  \bibinfo{author}{\bibfnamefont{I.}~\bibnamefont{Rotter}},
  \bibinfo{journal}{Phys. Rev. E} \textbf{\bibinfo{volume}{87}}
  (\bibinfo{year}{2013}).

\bibitem[{\citenamefont{Nataf et~al.}(2012)\citenamefont{Nataf, Baksic, and
  Ciuti}}]{Nataf2012Double}
\bibinfo{author}{\bibfnamefont{P.}~\bibnamefont{Nataf}},
  \bibinfo{author}{\bibfnamefont{A.}~\bibnamefont{Baksic}}, \bibnamefont{and}
  \bibinfo{author}{\bibfnamefont{C.}~\bibnamefont{Ciuti}},
  \bibinfo{journal}{Phys. Rev. A} \textbf{\bibinfo{volume}{86}},
  \bibinfo{pages}{013832} (\bibinfo{year}{2012}).

\bibitem[{\citenamefont{Baksic and Ciuti}(2014)}]{Baksic2014Controlling}
\bibinfo{author}{\bibfnamefont{A.}~\bibnamefont{Baksic}} \bibnamefont{and}
  \bibinfo{author}{\bibfnamefont{C.}~\bibnamefont{Ciuti}},
  \bibinfo{journal}{Phys. Rev. Lett.} \textbf{\bibinfo{volume}{112}},
  \bibinfo{pages}{173601} (\bibinfo{year}{2014}).

\bibitem[{\citenamefont{Lolli et~al.}(2015)\citenamefont{Lolli, Baksic, Nagy,
  Manucharyan, and Ciuti}}]{Lolli2015Ancillary}
\bibinfo{author}{\bibfnamefont{J.}~\bibnamefont{Lolli}},
  \bibinfo{author}{\bibfnamefont{A.}~\bibnamefont{Baksic}},
  \bibinfo{author}{\bibfnamefont{D.}~\bibnamefont{Nagy}},
  \bibinfo{author}{\bibfnamefont{V.~E.} \bibnamefont{Manucharyan}},
  \bibnamefont{and} \bibinfo{author}{\bibfnamefont{C.}~\bibnamefont{Ciuti}},
  \bibinfo{journal}{Phys. Rev. Lett.} \textbf{\bibinfo{volume}{114}},
  \bibinfo{pages}{183601} (\bibinfo{year}{2015}).

\bibitem[{\citenamefont{Paladino et~al.}(2014)\citenamefont{Paladino, Galperin,
  Falci, and Altshuler}}]{Paladino20141f}
\bibinfo{author}{\bibfnamefont{E.}~\bibnamefont{Paladino}},
  \bibinfo{author}{\bibfnamefont{Y.~.~M.} \bibnamefont{Galperin}},
  \bibinfo{author}{\bibfnamefont{G.}~\bibnamefont{Falci}}, \bibnamefont{and}
  \bibinfo{author}{\bibfnamefont{B.~.~L.} \bibnamefont{Altshuler}},
  \bibinfo{journal}{Rev. Mod. Phys.} \textbf{\bibinfo{volume}{86}},
  \bibinfo{pages}{361} (\bibinfo{year}{2014}).

\bibitem[{\citenamefont{Carusotto and Ciuti}(2013)}]{Carusotto2013Quantum}
\bibinfo{author}{\bibfnamefont{I.}~\bibnamefont{Carusotto}} \bibnamefont{and}
  \bibinfo{author}{\bibfnamefont{C.}~\bibnamefont{Ciuti}},
  \bibinfo{journal}{Rev. Mod. Phys.} \textbf{\bibinfo{volume}{85}},
  \bibinfo{pages}{299} (\bibinfo{year}{2013}).

\bibitem[{\citenamefont{Wachtel et~al.}(2016)\citenamefont{Wachtel, Sieberer,
  Diehl, and Altman}}]{Wachtel2016Electrodynamic}
\bibinfo{author}{\bibfnamefont{G.}~\bibnamefont{Wachtel}},
  \bibinfo{author}{\bibfnamefont{L.~M.} \bibnamefont{Sieberer}},
  \bibinfo{author}{\bibfnamefont{S.}~\bibnamefont{Diehl}}, \bibnamefont{and}
  \bibinfo{author}{\bibfnamefont{E.}~\bibnamefont{Altman}},
  \bibinfo{journal}{Phys. Rev. B} \textbf{\bibinfo{volume}{94}},
  \bibinfo{pages}{104520} (\bibinfo{year}{2016}).

\end{thebibliography}
\end{document}